%% file: alicepreprint_CDS_LMee-pp13TeV.tex
\documentclass[ALICE,manyauthors]{cernphprep}

\usepackage[comma,square,numbers,sort&compress]{natbib}
\usepackage[
        pdftitle={Dielectron production in pp collisions at sqrt(s) = 13 TeV},
        pdfauthor={ALICE Collaboration},
        pdfsubject={},
        bookmarksopen=false]{hyperref}
        
\usepackage{lineno}

\usepackage{booktabs}
\usepackage{amsmath}
\usepackage{amssymb} 
\usepackage{amsfonts}

\usepackage{color}
\usepackage{xspace}
\usepackage{siunitx}
\newcommand{\dd}{\ensuremath{\mathrm{d}}\xspace}

\newcommand{\dEdx}{\ensuremath{\dd E/\dd x}\xspace}
\newcommand{\eg}{e.\,g.\xspace}
\newcommand{\ie}{i.\,e.\xspace}

\newcommand{\pt}{\ensuremath{p_{\mathrm{T}}}\xspace}
\newcommand{\pte}{\ensuremath{p_{\mathrm{T,\,e}}}\xspace}
\newcommand{\ptee}{\ensuremath{p_{\mathrm{T,\,ee}}}\xspace}
\newcommand{\mt}{\ensuremath{m_{\mathrm{T}}}\xspace}
\newcommand{\mee}{\ensuremath{m_{\mathrm{ee}}}\xspace}
\newcommand{\etae}{\ensuremath{\eta_{\mathrm{e}}}\xspace}

\newcommand{\ee}{\ensuremath{\mathrm{e}^+\mathrm{e}^-}\xspace}

\newcommand{\bbbar}{\ensuremath{\mathrm{b}\overline{\mathrm{b}}}\xspace}
\newcommand{\ccbar}{\ensuremath{\mathrm{c}\overline{\mathrm{c}}}\xspace}

\newcommand{\Jpsi}{\ensuremath{\mathrm{J}\hspace{-.08em}/\psi}\xspace}
\newcommand{\psiP}{\ensuremath{\psi\mathrm{(2S)}}\xspace}

\newcommand{\D}{\ensuremath{\mathrm{D}}\xspace}

\newcommand{\GeV}{\ensuremath{~\si{\giga\electronvolt}}\xspace}
\newcommand{\TeV}{\ensuremath{~\si{\tera\electronvolt}}\xspace}

\newcommand{\GeVc}{\ensuremath{~\si{\giga\electronvolt\per\mathit{c}}}\xspace}

\newcommand{\GeVcc}{\ensuremath{~\si{\giga\electronvolt\per\mathit{c}^2}}\xspace}

\newcommand{\pp}{{\ensuremath{\mathrm{pp}}}\xspace}

\newcommand{\dAu}{\ensuremath{\text{d--Au}}\xspace}

\newcommand{\sqrts}{\ensuremath{\sqrt{s}}\xspace}
\newcommand{\sqrtsnn}{\ensuremath{\sqrt{s_{_{\mathrm{NN}}}}}\xspace}

\newcommand{\mb}{\ensuremath{~\si{\milli\barn}}\xspace}
\newcommand{\mub}{\ensuremath{~\si{\micro\barn}}\xspace}

\newcommand{\ccFONLL}{1296}
\newcommand{\ccFONLLehigh}{172}
\newcommand{\ccFONLLelow}{162}
\newcommand{\bbFONLL}{68}
\newcommand{\bbFONLLehigh}{15}
\newcommand{\bbFONLLelow}{16}

\newcommand{\ccPYTHIA}{974}
\newcommand{\ccPYTHIAstat}{138}
\newcommand{\ccPYTHIAsyst}{140}
\newcommand{\bbPYTHIA}{79}
\newcommand{\bbPYTHIAstat}{14}
\newcommand{\bbPYTHIAsyst}{11}
\newcommand{\ccPOWHEG}{1417}
\newcommand{\ccPOWHEGstat}{184}
\newcommand{\ccPOWHEGsyst}{204}
\newcommand{\bbPOWHEG}{48}
\newcommand{\bbPOWHEGstat}{14}
\newcommand{\bbPOWHEGsyst}{7}
\newcommand{\ccHMPYTHIA}{4.14}
\newcommand{\ccHMPYTHIAstat}{0.67}
\newcommand{\ccHMPYTHIAsyst}{0.66}
\newcommand{\bbHMPYTHIA}{0.29}
\newcommand{\bbHMPYTHIAstat}{0.07}
\newcommand{\bbHMPYTHIAsyst}{0.05}
\newcommand{\ccHMPOWHEG}{5.95}
\newcommand{\ccHMPOWHEGstat}{0.91}
\newcommand{\ccHMPOWHEGsyst}{0.95}
\newcommand{\bbHMPOWHEG}{0.17}
\newcommand{\bbHMPOWHEGstat}{0.07}
\newcommand{\bbHMPOWHEGsyst}{0.03}

\newcommand{\DPHonetwo}{0.057}
\newcommand{\DPHtwothree}{0.072}
\newcommand{\DPHthreesix}{0.023}
\newcommand{\DPHHMonetwo}{0.060}
\newcommand{\DPHHMtwothree}{0.083}
\newcommand{\DPHHMthreesix}{0.055}

\newcommand{\nbinv} {\ensuremath{~\si{\per\nano\barn}}\xspace}
\newcommand{\pbinv} {\ensuremath{~\si{\per\pico\barn}}\xspace}

\newcommand{\Lumi}{\ensuremath{\mathcal{L}}\xspace}

\providecommand{\PYTHIA} {{\textsc{Pythia}}\xspace}

\providecommand{\GEANT} {{\textsc{Geant}}\xspace}

\providecommand{\POWHEG} {{\textsc{Powheg}}\xspace}

 \graphicspath{{Figures/}}
\begin{document}

\begin{titlepage}
 \PHyear{2018}
 \PHnumber{122}      
 \PHdate{9 May}  

\title{Dielectron and heavy-quark production in inelastic and high-multiplicity proton--proton collisions at $\mathbf{\sqrt{{\textit s}} = 13}$~TeV}
\ShortTitle{Dielectron production in \pp collisions at $\sqrts = 13\TeV$} 

\Collaboration{ALICE Collaboration\thanks{See Appendix~\ref{app:collab} for the list of collaboration members}}
\ShortAuthor{ALICE Collaboration} 

\begin{abstract}
  The measurement of dielectron production is presented as a function of
  invariant mass and transverse momentum ($p_{\rm T}$) at midrapidity
  ($|y_{\rm e}|<0.8$) in proton-proton (pp) collisions at a centre-of-mass
  energy of $\sqrt{s}=13$~TeV. The contributions from light-hadron decays are
  calculated from their measured cross sections in pp collisions at
  $\sqrt{s}=7$~TeV or 13~TeV. The remaining continuum stems from correlated
  semileptonic decays of heavy-flavour hadrons. Fitting the data with templates
  from two different MC event generators, PYTHIA and POWHEG, the charm and
  beauty cross sections at midrapidity are extracted for the first time at this
  collision energy:
  ${\rm d}\sigma_{\rm c\bar{c}}/{\rm
    d}y|_{y=0}=974\pm138(\rm{stat.})\pm140(\rm{syst.})\pm214(\rm{BR})~\mu{\rm b}$ and
  ${\rm d}\sigma_{\rm b\bar{b}}/{\rm
    d}y|_{y=0}=79\pm14(\rm{stat.})\pm11(\rm{syst.})\pm5(\rm{BR})~\mu{\rm b}$ using PYTHIA
  simulations and
  ${\rm d}\sigma_{\rm c\bar{c}}/{\rm
    d}y|_{y=0}=1417\pm184(\rm{stat.})\pm204(\rm{syst.})\pm312(\rm{BR})~\mu{\rm b}$ and
  ${\rm d}\sigma_{\rm b\bar{b}}/{\rm
    d}y|_{y=0}=48\pm14(\rm{stat.})\pm7(\rm{syst.})\pm3(\rm{BR})~\mu{\rm b}$ for POWHEG. These
  values, whose uncertainties are fully correlated between the two generators,
  are consistent with extrapolations from lower energies. The different results
  obtained with POWHEG and PYTHIA imply different kinematic correlations of the
  heavy-quark pairs in these two generators. Furthermore, comparisons of
  dielectron spectra in inelastic events and in events collected with a trigger
  on high charged-particle multiplicities are presented in various $p_{\rm T}$
  intervals. The differences are consistent with the already measured scaling of
  light-hadron and open-charm production at high charged-particle multiplicity
  as a function of $p_{\rm T}$. Upper limits for the contribution of virtual
  direct photons are extracted at 90\% confidence level and found to be in
  agreement with pQCD calculations.
\end{abstract}
\end{titlepage}
\setcounter{page}{2}

\maketitle 

\section{Introduction}
\label{sec:intro}
Heavy-flavour quarks (charm and beauty) are copiously produced by inelastic
partonic scatterings in high-energy proton--proton (\pp) collisions at the CERN
Large Hadron Collider (LHC). Their large masses ($m_{\rm Q}$) make it possible
to calculate their production cross sections with perturbative quantum
chromodynamics (pQCD)~\cite{Cacciari:1998it,Cacciari:2001td,Cacciari:2012ny}.
Hence, experimental measurements of heavy-quark production provide an excellent
test of pQCD in this energy regime. Flavour conservation allows heavy quarks to
be only produced in pairs. Charm hadrons and their decay products reflect the
initial angular correlation of the heavy-quark pairs, whereas in the case of
decays of beauty hadrons the correlation is weakened due to their large masses.
The contribution from the simultaneous semileptonic decays of the corresponding
heavy-flavour hadron pairs dominates the dilepton yield in the intermediate mass
region (IMR) $1<m_{\ell\ell}<3\GeVcc$. Hence, dielectron measurements can be
used to study charm and beauty production.

The ALICE Collaboration has reported charm and beauty production cross sections
measurements at midrapidity ($\vert y\vert<0.5$) in \pp collisions at
centre-of-mass energies of $\sqrts = 2.76$ and
$7\TeV$~\cite{Abelev:2012vra,Abelev:2014hla,Abelev:2014gla,Adam:2016ich,Acharya:2017jgo,Abelev:2012gx,Abelev:2012sca}.
The charm measurement at $\sqrts = 7\TeV$ is complemented by ATLAS data
extending to higher transverse momentum (\pt) and $|y|<2.1$~\cite{Aad:2015zix}.
Furthermore, the CMS Collaboration has provided a variety of charm and bottom
measurements at midrapidity at $\sqrts = 2.76$, 5 and
7\TeV~\cite{Khachatryan:2011mk,Khachatryan:2011hf,Chatrchyan:2011pw,Chatrchyan:2011kc,Chatrchyan:2012np,Chatrchyan:2012hw,Sirunyan:2017oug,Sirunyan:2017xss,Sirunyan:2017isk}.
At forward rapidity ($2<y<5$), the LHCb Collaboration has measured charm and
beauty production cross sections in \pp collisions at $\sqrts = 5$, 7, 8 and
13\TeV~\cite{Aaij:2013yaa,Aaij:2015bpa,Aaij:2016jht,Aaij:2016avz}. These results
are generally in good agreement with pQCD calculations at next-to-leading order
(NLO) in the strong coupling constant ($\alpha_{\rm s}$) with all-order
resummation of the logarithms of $\pt/m_{\rm Q}$
(FONLL)~\cite{Cacciari:1998it,Cacciari:2001td,Cacciari:2012ny}. Though the
measured charm production cross sections consistently lie on the upper edge of
the systematic uncertainties of the theory calculations. Recently, the ALICE
Collaboration has measured the charm and beauty production cross sections in \pp
collisions at $\sqrts = 7\TeV$ using electron--positron pairs (dielectrons) from
correlated semileptonic decays of heavy-flavour
hadrons~\cite{Acharya:2018ohw}. Such an approach was first performed by the
PHENIX Collaboration in \pp and \dAu collisions at $\sqrtsnn = 200\GeV$ at the
Relativistic Heavy Ion Collider
(RHIC)~\cite{Adare:2008ac,Adare:2014iwg,Adare:2017caq}. These measurements have
the advantage that they probe the full \pt range of heavy-quark pairs and
contain complementary information about the initial correlation of charm quarks,
\ie\ the underlying production mechanism, which is not accessible in
conventional single heavy-flavour measurements.

The measurement of direct photons, \ie those produced in hard scatterings
between incoming partons in hadronic collisions, provides another important test
of pQCD. Furthermore, at $\pt < 3\GeVc$, where the applicability of perturbation
theory may be questionable, experimental data of direct-photon production in \pp
collisions serve as a crucial reference to establish the presence of thermal
radiation from the hot and dense medium created in heavy-ion
collisions~\cite{Adare:2008ab,STAR:2016use,Adam:2015lda,Acharya:2018dqe}. The
measurement of real (massless) direct photons at low \pt is challenging because
of the large background of hadron decay photons. This background can be largely
reduced by measuring the contribution of virtual direct photons, \ie\ direct \ee
pairs, to the dielectron invariant-mass spectrum above the $\pi^0$
mass~\cite{Adare:2008ab,STAR:2016use}.

Proton--proton collisions in which a large number of charged particles are
produced have recently attracted the interest of the heavy-ion
community~\cite{Loizides:2016tew,Nagle:2018nvi}. These events exhibit features
that are similar to those observed in heavy-ion collisions, \eg\ collective
effects, such as long-range angular
correlations~\cite{Khachatryan:2010gv,Khachatryan:2016txc,Abelev:2013sqa,Aad:2015gqa,Aaboud:2016yar,Aaboud:2017acw}
or enhanced strangeness production~\cite{ALICE:2017jyt}. Charged-hadron \pt
spectra in \pp collisions at $\sqrts = 13\TeV$ show a hardening with increasing
multiplicity, an effect that arises naturally from jets~\cite{Adam:2015pza}.
Also, heavy-quark production is found to scale faster than linearly with the
charged-particle multiplicity in \pp collisions at
$\sqrts = 7\TeV$~\cite{Adam:2015ota,Abelev:2012rz}. This motivates the study of
dielectron production in high-multiplicity \pp collisions. In the low mass
region ($\mee<1\GeVcc$), dielectron measurements provide further insight into
possible modifications of the light vector and pseudo-scalar meson production
via their resonance and/or Dalitz decays, whereas in the IMR they allow for
complementary studies of the heavy-flavour production. At LHC energies, the
contribution from open charm already dominates the dielectron continuum at
$\mee\approx0.5\GeVcc$. Moreover, if a thermalised system were created in such
high-multiplicity \pp collisions, a signal of thermal (virtual) photons should
be present.

In this letter, first results of charm and beauty production cross sections at
midrapidity in inelastic (INEL) and high-multiplicity (HM) \pp collisions at
$\sqrts = 13\TeV$ are reported. The paper is organised as follows: the ALICE
apparatus and the data samples used in the analysis are described in
Section~\ref{sec:detector}, the data analysis is discussed in
Section~\ref{sec:signal}, Section~\ref{sec:cocktail} introduces the cocktail of
known hadronic sources, and the results are presented and discussed in
Section~\ref{sec:results}.

\section{The ALICE detector and data samples}
\label{sec:detector}
A detailed description of the ALICE apparatus and its performance can be found
in~\cite{Carminati:2004fp,Alessandro:2006yt,Aamodt:2008zz,Abelev:2014ffa}. The
detectors used in this analysis are briefly described below.

Trajectories of charged particles are reconstructed in the ALICE central barrel
with the Inner Tracking System (ITS) and the Time Projection Chamber (TPC) that
reside within a solenoid, which provides a homogeneous magnetic field of
\SI{0.5}{\tesla} along the beam direction. The ITS consists of six cylindrical
layers of silicon detectors, with radial distances from the beam axis between
$\SI{3.9}{\centi\metre}$ and $\SI{43}{\centi\metre}$. The two innermost layers
are equipped with Silicon Pixel Detectors (SPD), the two intermediate layers are
composed of Silicon Drift Detectors, and the two outermost layers are made of
Silicon Strip Detectors. The TPC, main tracking device in the ALICE central
barrel, is a $\SI{5}{\metre}$ long cylindrical gaseous detector extending from
$\SI{85}{\centi\metre}$ to $\SI{247}{\centi\metre}$ in radial direction. It
provides up to 159 spacial points per track for charged-particle reconstruction
and particle identification (PID) through the measurement of the specific
ionisation energy loss \dEdx in the gas volume.

The PID is complemented by the Time-Of-Flight (TOF) system located at a radial
distance of $\SI{3.7}{\metre}$ from the nominal interaction point. It measures
the arrival time of particles relative to the event collision time provided by
the TOF detector itself or by the T0 detectors, two arrays of Cherenkov-counters
located at forward rapidities~\cite{Adam:2016ilk}.

Collision events are triggered by the V0 detector that comprises two plastic
scintillator arrays placed on both sides of the interaction point at
pseudorapidities $2.8 < \eta < 5.1$ and $-3.7 < \eta < -1.7$. The V0 is also
used to reject background events like beam-gas interactions, collisions with
de-bunched protons or with mechanical structures of the beam line.

The data samples used in this letter were recorded with ALICE in 2016 during the
LHC \pp run at $\sqrts = 13\TeV$. For the minimum-bias event trigger that is
used to define the data sample for the analysis of inelastic \pp collisions,
coincident signals in both V0 scintillators are required to be synchronous with
the beam crossing time defined by the LHC clock. Events with high
charged-particle multiplicities are triggered on by additionally requiring the
total signal amplitude measured in the V0 detector to exceed a certain
threshold. At the analysis level, the 0.036 percentile of inelastic events with
the highest V0 multiplicity (V0M) is selected to define the high-multiplicity
event class. This value is low enough to avoid inefficiencies due to trigger
threshold variations during data taking. Track segments reconstructed in the SPD
are extrapolated back to the beam line to define the interaction vertex. Events
with multiple vertices identified with the SPD are tagged as pile-up and removed
from the analysis~\cite{Abelev:2014ffa}. The vertex information may be improved
based on the information provided by tracks reconstructed in the ITS and TPC. To
assure a uniform detector coverage within $\vert\eta\vert<0.8$, the vertex
position along the beam direction is restricted to $\pm\SI{10}{\centi\metre}$
around the nominal interaction point. A total of $455\times 10^{6}$ minimum-bias
(MB) \pp events and $79.2\times 10^{6}$ high-multiplicity \pp events are
considered for further analysis, which corresponds to an integrated luminosity
of $\Lumi_{\mathrm{int}}^{\mathrm{MB}} = 7.87\pm0.40\nbinv$ and
$\Lumi_{\mathrm{int}}^{\mathrm{HM}} = 2.79\pm0.15\pbinv$, respectively. The
luminosity determination is based on the visible cross section for the V0-based
minimum-bias trigger, measured in a van der Meer scan carried out in
2015~\cite{ALICE-PUBLIC-2016-002}. A conservative uncertainty of 5\% is assigned
to this measurement, to account for possible variations of the cross section
between the two data-taking periods.

\section{Data analysis}
\label{sec:signal}
Electron\footnote{The term `electron' is used for both electrons and positrons
  if not stated otherwise.} candidates are selected from charged-particle tracks
reconstructed in the ITS and TPC in the kinematic range $\vert\etae\vert < 0.8$
and $\pte > 0.2\GeVc$. Basic track quality criteria are applied, \eg\ a
sufficient number of space points measured in the TPC and ITS as well as a good
track fit.
The contribution from secondary tracks is reduced by requiring a maximum
distance of closest approach (DCA) to the primary vertex in the transverse plane
(DCA$_{\rm xy} < \SI{1.0}{\centi\metre}$) and in the longitudinal direction
(DCA$_{\rm z} < \SI{3.0}{\centi\metre}$). To further suppress the contribution
from photon conversions in the detector material, electron candidates are
required to have a hit in the first SPD layer and no ITS clusters shared with
other reconstructed tracks.

The electron identification is based on the complementary information provided
by the TPC and TOF. The detector PID response, $n(\sigma_{i}^{\rm DET})$, is
expressed in terms of the deviation between the measured and expected value of
the specific ionisation energy loss in the TPC or time-of-flight in the TOF for
a given particle hypothesis $i$ and momentum, normalised by the detector
resolution $(\sigma^{\rm DET})$. In the TPC, electrons are selected in the range
$\left\vert n\left(\sigma_{\rm e}^{\rm TPC}\right) \right\vert < 3$ and pions
are rejected by requiring $n\left(\sigma_{\rm \pi}^{\rm TPC}\right) > 4$.
Furthermore, kaons and protons are rejected with
$\left\vert n\left(\sigma_{\rm K}^{\rm TPC}\right) \right\vert > 4$ and
$\left\vert n\left(\sigma_{\rm p}^{\rm TPC}\right) \right\vert > 4$, unless the
candidate is positively identified as an electron in the TOF, \ie\ fulfilling
$\left\vert n\left(\sigma_{\rm e}^{\rm TOF}\right) \right\vert < 3$. For
particles that are outside
$\left\vert n\left(\sigma_{\rm K}^{\rm TPC}\right) \right\vert < 4$ and
$\left\vert n\left(\sigma_{\rm p}^{\rm TPC}\right) \right\vert < 4$ the TOF
information is only used to select electron candidates with
$\left\vert n\left(\sigma_{\rm e}^{\rm TOF}\right) \right\vert < 3$ if the track
has an associated hit in the TOF detector.

\begin{figure}[ht]
  \centering
  \includegraphics[width=0.49\linewidth]{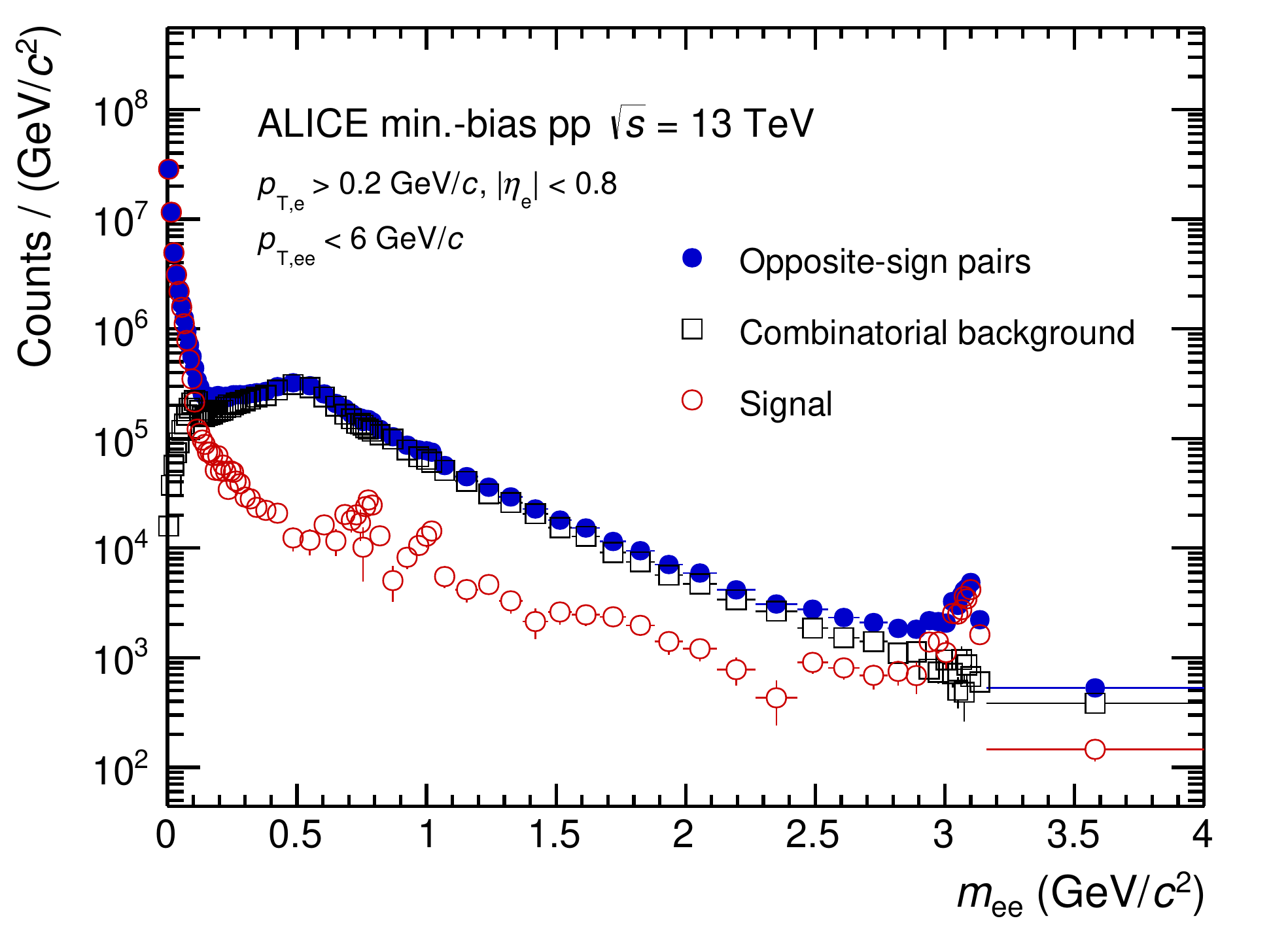}
  \hfill
  \includegraphics[width=0.49\linewidth]{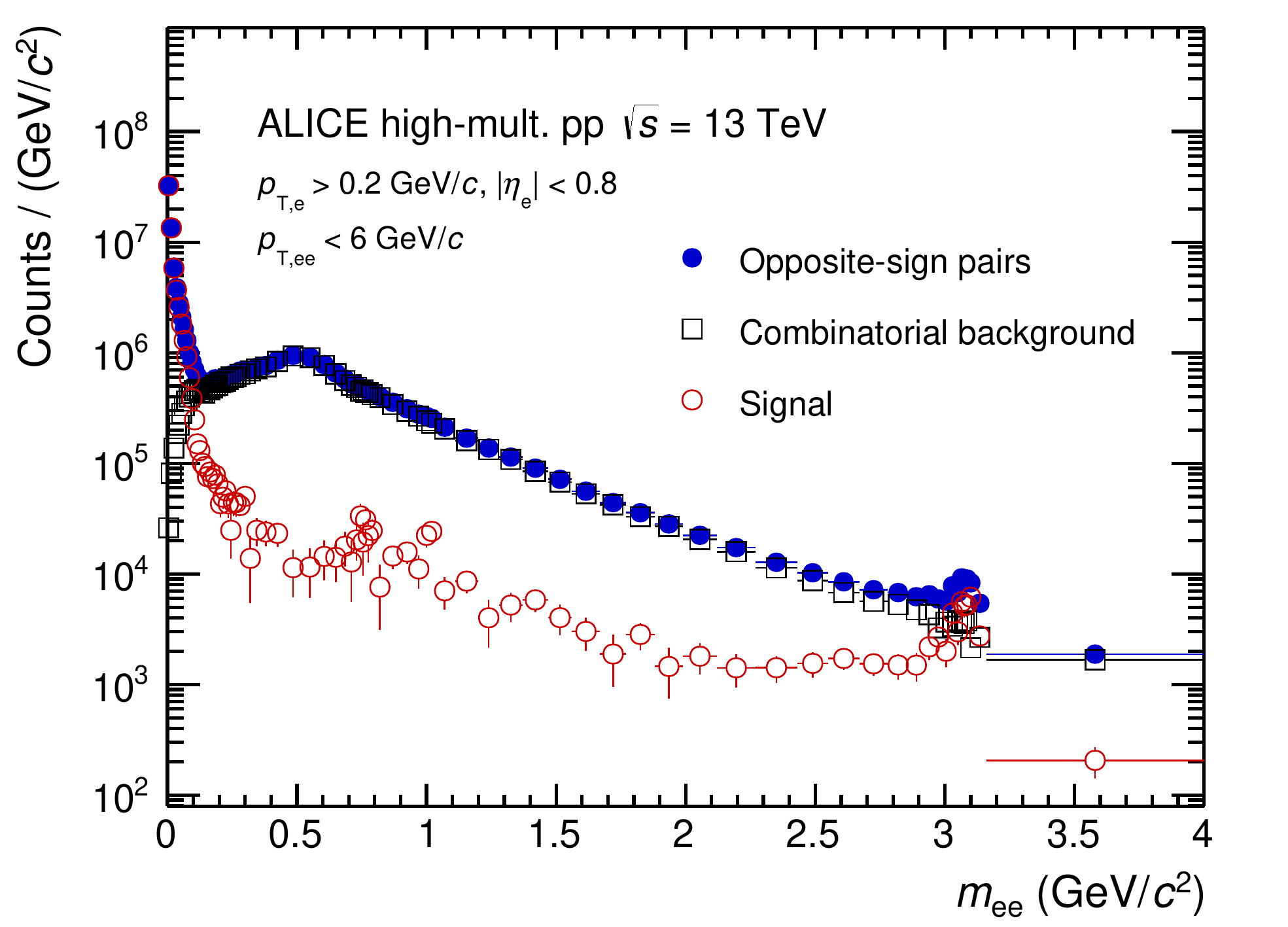}
  \caption{Opposite-sign spectrum $N_{+-}$, the combinatorial background $B$ and
    the signal $S$ in minimum-bias (left) and high-multiplicity (right) events.
    Only statistical uncertainties are shown.}
  \label{fig:ULS_LS_S_nocut}
\end{figure}

Since experimentally the origin of each electron or positron is unknown, all
electron candidates are paired considering combinations with opposite $(N_{+-})$
but also same-sign charge $(N_{\pm\pm})$. Most of the electron pairs arise from
the combination of two electrons originating from different mother particles.
These pairs give rise to the combinatorial background $B$ that is estimated via
the geometric mean of same-sign pairs $\sqrt{N_{++} N_{--}}$ within the same
event. Opposite- and same-sign pairs include correlated background, \eg\
originating from $\pi^0$ decays with two \ee pairs in the final state
($\pi^0 \rightarrow \gamma^{(\ast)} \gamma^{(\ast)}\rightarrow \ee \ee$), which
includes decay channels with real photons and their subsequent conversion in
detector material. Such processes lead to opposite and same-sign pairs at equal
rate. The background estimate needs to be corrected for the different detector
acceptance of opposite and same-sign pairs. This correction factor is determined
by dividing the yields of uncorrelated opposite $(M_{+-})$ and same-sign pairs
$(M_{\pm\pm})$ in mixed events: $R = M_{+-} / (2\sqrt{M_{++} M_{--}})$. The
dielectron signal is then obtained as
$S = N_{+-} - B = N_{+-} - 2R\sqrt{N_{++} N_{--}}$. The signal $S$ is shown
together with the opposite-sign spectrum $N_{+-}$ and the combinatorial
background $B$ in Fig.~\ref{fig:ULS_LS_S_nocut} for minimum-bias and
high-multiplicity events. In the mass interval $0.2<\mee<3\GeVcc$, the
signal-to-background ratio varies in MB events between 0.3 and 0.04 with a
minimum around $\mee \approx 0.5\GeVcc$ and is roughly constant at 0.2 in the
IMR~\cite{ALICE-PUBLIC-2018-009}. In HM events, the minimum reaches 0.01 and is
about 0.08 in the IMR.

Electron--positron pairs from photon conversion in the detector material,
contributing to the low mass spectrum below 0.14\GeVcc, are removed by using
their distinct orientation relative to the magnetic
field~\cite{Acharya:2018ohw}.

The data are corrected for the reconstruction efficiencies using detailed Monte
Carlo (MC) simulations. For this, \pp events are generated with the Monash 2013
tune of \PYTHIA~8~\cite{Skands:2014pea} for light-hadron decays and the Perugia
2011 tune of \PYTHIA~6.4 for heavy-flavour decays~\cite{Sjostrand:2006za} and
the resulting particles are propagated through a detector simulation using
\GEANT~3~\cite{Brun:118715}. The choice of the different \PYTHIA versions is
motivated by the fact that the Perugia 2011 tune describes reasonably well the
transverse momentum spectra of heavy-flavour hadrons while the Monash 2013 tune
reproduces many of the relevant light-hadron
multiplicities~\cite{Acharya:2018qnp,ALICE-PUBLIC-2018-004}. The signal reconstruction
efficiencies were studied as a function of \mee and pair transverse momentum
\ptee separately for the different \ee sources: resonance and Dalitz decays of
relevant mesons as well as correlated semileptonic decays of charm and beauty
hadrons. The total signal reconstruction efficiency is obtained by weighting the
efficiency of each dielectron source by its expected contribution and is found
to be about $20\%$ in $0.7<\mee<1.2\GeVcc$ and approaches 30\% at lower and
higher masses.

Different aspects of the analysis are considered as possible sources of
systematic uncertainties, which are summarised in Table~\ref{tab:SystUncert}.
The systematic uncertainties due to the track reconstruction are estimated by
comparing the efficiency of the ITS--TPC matching, the requirement of a hit in
the first SPD layer, and the requirement of no shared ITS clusters in MC
simulations and data. The residual disagreements between data and MC add to a
6.5\% uncertainty on the single track level, which leads to a 13\% uncertainty
for pairs. The MC simulations were also checked to reproduce all details of the
PID selection within a systematic uncertainty of 2\% for \ee pairs. The purity
of the electron sample is estimated to be ${>}93\%$ over the relevant \pt range,
with a \pt-integrated hadron contamination of about 4\%. The resulting hadron
contamination on the dielectron signal is found to be negligible. For
$\mee<0.14\GeVcc$, a 2\% uncertainty on the conversion rejection was estimated
from the yield change when tightening the selection to reject photon
conversions. A 2\% uncertainty on the signal yield due to the correction factor
$R$ is obtained by repeating the event mixing in different event classes,
defined by the position of the reconstructed primary vertex and by the
charged-particle multiplicity. The efficiency of the minimum-bias trigger to
select inelastic events with an \ee pair in the ALICE acceptance
($\vert \etae \vert<0.8$ and $\pte>0.2\GeVc$) is estimated to be $(99\pm1)\%$
from the Monash 2013 tune of \PYTHIA~8. This and the luminosity uncertainty of
5\%~\cite{ALICE-PUBLIC-2016-002} are global uncertainties, which are not
included in the point-to-point uncertainties. No significant variation of
systematic uncertainties on mass or $\ptee$ is observed in the analysis, and the
same total uncertainty of 14\% is assigned as point-to-point correlated
uncertainties on the differential dielectron cross section in inelastic \pp
collisions.

The analysis of the high-multiplicity data has additional systematic
uncertainties. First, no dedicated high-multiplicity MC simulation was
performed. In such events the vertex distribution is biased more than in MB
events by the asymmetric pseudorapidity coverage of the two V0 detectors. The
change of the detector acceptance with vertex position could lead to a
difference in the number of reconstructed electrons of up to 3\%, which results
in an uncertainty of 6\% for \ee pairs. Second, a possible multiplicity
dependence of the reconstruction and PID efficiency is covered by an uncertainty
of 6\%~\cite{Abelev:2013haa}. Added in quadrature, this amounts to a total
uncertainty of 15\%.

\begin{table}[ht]
  \centering
  \begin{tabular}{l r r}
    \toprule
    Source & Minimum bias & High multiplicity \\
    \midrule
    Track reconstruction & 13\% & 13\%\\
    Electron identification & 2\% & 2\% \\
    Conversion rejection ($\mee < 0.14\GeVcc$) & 2\%  & 2\% \\
    Acceptance correction factor ($R$) & 2\% & 2\% \\
    Vertex distribution bias & -- & 6\%\\
    Multiplicity dependence of tracking and PID & -- & 6\% \\
    \midrule
    Total & 14\% & 15\% \\
    \bottomrule
  \end{tabular}
  \caption{Sources of systematic uncertainties.}
  \label{tab:SystUncert}
\end{table}

\section{Cocktail of known hadronic sources}
\label{sec:cocktail}
The dielectron spectrum measured in \pp collisions at $\sqrts = 13\TeV$ is
compared with the expectations from all known hadron sources, \ie\ the hadronic
cocktail, contributing to the dielectron spectrum in the ALICE central barrel
acceptance ($\vert\etae\vert<0.8$ and $\pte>0.2\GeVc$). A fast MC simulation is
used to estimate the contribution from $\pi^0$, $\eta$, $\eta'$, $\rho$,
$\omega$ and $\phi$ decays in \pp collisions, as detailed
in~\cite{Acharya:2018ohw}.

Following the approach outlined in~\cite{Acharya:2018qsh}, the pion \pt-spectrum
at $\sqrts = 13\TeV$ is approximated by scaling the \pt-spectrum of charged
hadrons~\cite{Adam:2015pza} by the pion-to-hadron ratio measured at
$\sqrts = 7\TeV$~\cite{Abelev:2013ala,Adam:2016dau}. The difference with respect
to the same procedure based on the pion-to-hadron ratio measured at
$\sqrts = 2.76\TeV$~\cite{Abelev:2013ala,Abelev:2014laa} is smaller than 1\% at
low \pt and reaches 5\% at high \pt. The charged hadron \pt-spectra at
$\sqrts = 13\TeV$ are normalised to INEL${>}0$ events, \ie\ inelastic collisions
that produce at least one charged particle in $|\eta|<1$, rather than INEL
events. This is corrected taking the 21\% difference in the \pt integrated
$\dd N_{\rm ch}/\dd \eta$ values for these two event
classes~\cite{Adam:2015pza}. A conservative uncertainty of 10\% is assigned on
this extrapolation.

A fit of the obtained charged-pion \pt-spectrum with a modified Hagedorn
function is then taken as proxy for the neutral-pion \pt-distribution. The
simulated cross section per unit rapidity of the $\pi^0$ is
$\dd \sigma/\dd y\vert_{y=0} = 155.2\mb$. For the $\eta$ meson a fit of the
measured $\eta/\pi^0$ ratio in \pp collisions at $\sqrts = 7\TeV$ is
used~\cite{Abelev:2012cn}. The Monash 2013 tune of \PYTHIA~8 describes the
$\rho/\pi^0$ and $\omega/\pi^0$ ratios measured in \pp collisions at
$\sqrts = 2.76$ and 7\TeV,
respectively~\cite{Acharya:2018qnp,ALICE-PUBLIC-2018-004}. Therefore, MC
simulations obtained with this tune at $\sqrts = 13\TeV$ are used to obtain the
$\rho/\pi^0$ and $\omega/\pi^0$ ratios. Based on the $\eta/\pi^0$, $\rho/\pi^0$
and $\omega/\pi^0$ data, the ratios at high \pt are $0.5 \pm 0.1$, $1.0 \pm 0.2$
and $0.85 \pm 0.17$, respectively. The $\eta^{\prime}$ and $\phi$ mesons are
generated assuming \mt scaling, replacing \pt with
$\sqrt{m^2 - m_{\pi}^2 + (\pt/c)^2}$~\cite{Altenkamper:2017qot}. For the \mt
scaling, particle yields are normalised at high \pt relative to the $\pi^0$
yield: $0.40 \pm 0.08$ for $\eta^{\prime}$ (from \PYTHIA~6 calculations) and
$0.13 \pm 0.04$ for $\phi$~\cite{Abelev:2012hy}. The detector response,
including momentum and angular resolutions, as well as Bremsstrahlung effects
obtained from full MC simulations, is applied to the decay electrons as a
function of $\pte$, $\eta_{\rm e}$ and the azimuth $\varphi_{\rm e}$. This
results in a mass resolution of approximately 1\%. The following sources of
systematic uncertainties were evaluated: the input parameterisations of the
measured spectra as a function of \pt ($\pi^{\pm}$, $\eta/\pi^0$ and
$\omega/\pi^0$), the branching fractions of all included decay modes, the \mt
scaling parameters and the resolution smearing. For the high-multiplicity
cocktail, the input hadron \pt-distributions are adjusted according to the
measured modifications of the charged-hadron \pt spectra~\cite{Adam:2015pza}.
The uncertainties of the cocktail from light-hadron decays are about $\pm15\%$,
reaching up to $+50\%$ in the region dominated by the $\eta$ meson due to
uncertainties in the extrapolation to low \pt. The multiplicity dependence has
an uncertainty that varies between about 12\% at low \pt and 40\% at high \pt.

The Perugia~2011 tune of \PYTHIA~6.4, which includes NLO parton showering
processes, is used to estimate the contributions of correlated semileptonic
decays of open charm and beauty hadrons~\cite{Skands:2010ak,Sjostrand:2006za}.
As an alternative, the NLO event generator \POWHEG is also
considered~\cite{Nason:2004rx,Frixione:2007nw,Frixione:2007vw,Alioli:2010xd}.
The resulting same-sign spectrum is subtracted from the opposite-sign
distribution as in the data analysis. Detector effects are implemented as for
the light-hadron cocktail. The spectra are normalised to cross sections at
midrapidity that are based on
FONLL~\cite{Cacciari:1998it,Cacciari:2001td,Cacciari:2012ny} extrapolations of
the ALICE measurements at
7\TeV~\cite{Acharya:2017jgo,Abelev:2012gx,Abelev:2012sca}.
Following the description in~\cite{Averbeck:2011ga}, this leads to cross
sections per unit rapidity of
$\dd\sigma_{\ccbar}/\dd y\vert_{y=0} = \ccFONLL^{+\ccFONLLehigh}_{-\ccFONLLelow}\mub$ and
$\dd\sigma_{\bbbar}/\dd y\vert_{y=0} = \bbFONLL^{+\bbFONLLehigh}_{-\bbFONLLelow}\mub$ at $\sqrts = 13\TeV$.
The quoted uncertainties take into account both the measured uncertainty and the
FONLL extrapolation uncertainties. The latter (dominated by scale uncertainties
but also including PDF and mass uncertainties) are considered to be fully
correlated between the two energies~\cite{Cacciari:2015fta}.
For the high-multiplicity cocktail, the open charm contribution is weighted as a
function of \pt according to the measured enhancement of \D mesons with
$\pt>1\GeVc$ at $\sqrts = 7\TeV$~\cite{Adam:2015ota}. The same weights are
applied to the open beauty contribution as no significant difference between the
production of \D mesons and \Jpsi from beauty-hadron decays is
observed~\cite{Adam:2015ota}. For electrons originating from charm or beauty
hadrons with $\pt<1\GeVc$, the same weight as for $1<\pt<2\GeVc$ is assumed in
the absence of a measurement. This leads to an uncertainty on the multiplicity
dependence of about 40\% at low \pt decreasing to 20\% at high \pt.

The \Jpsi contribution is simulated with \PYTHIA~6.4 and normalised to the cross
section at $\sqrts = 13\TeV$, extrapolated with FONLL~\cite{Abelev:2012gx} from
the measurement at $\sqrts = 7\TeV$ by the ALICE
Collaboration~\cite{Aamodt:2011gj}. In the high-multiplicity cocktail, the \Jpsi
is scaled according to a dedicated, \pt-integrated
measurement~\cite{Abelev:2012rz}. The \psiP contribution is normalised to the
\Jpsi based on a cross section ratio of
$\sigma_{\psiP\rightarrow\ee}\slash\sigma_{\Jpsi\rightarrow\ee} =
(1.59\pm0.17)\%$~\cite{Adam:2015rta}.

\section{Results}
\label{sec:results}
\begin{figure}[ht]
  \centering
  \includegraphics[width=0.49\linewidth]{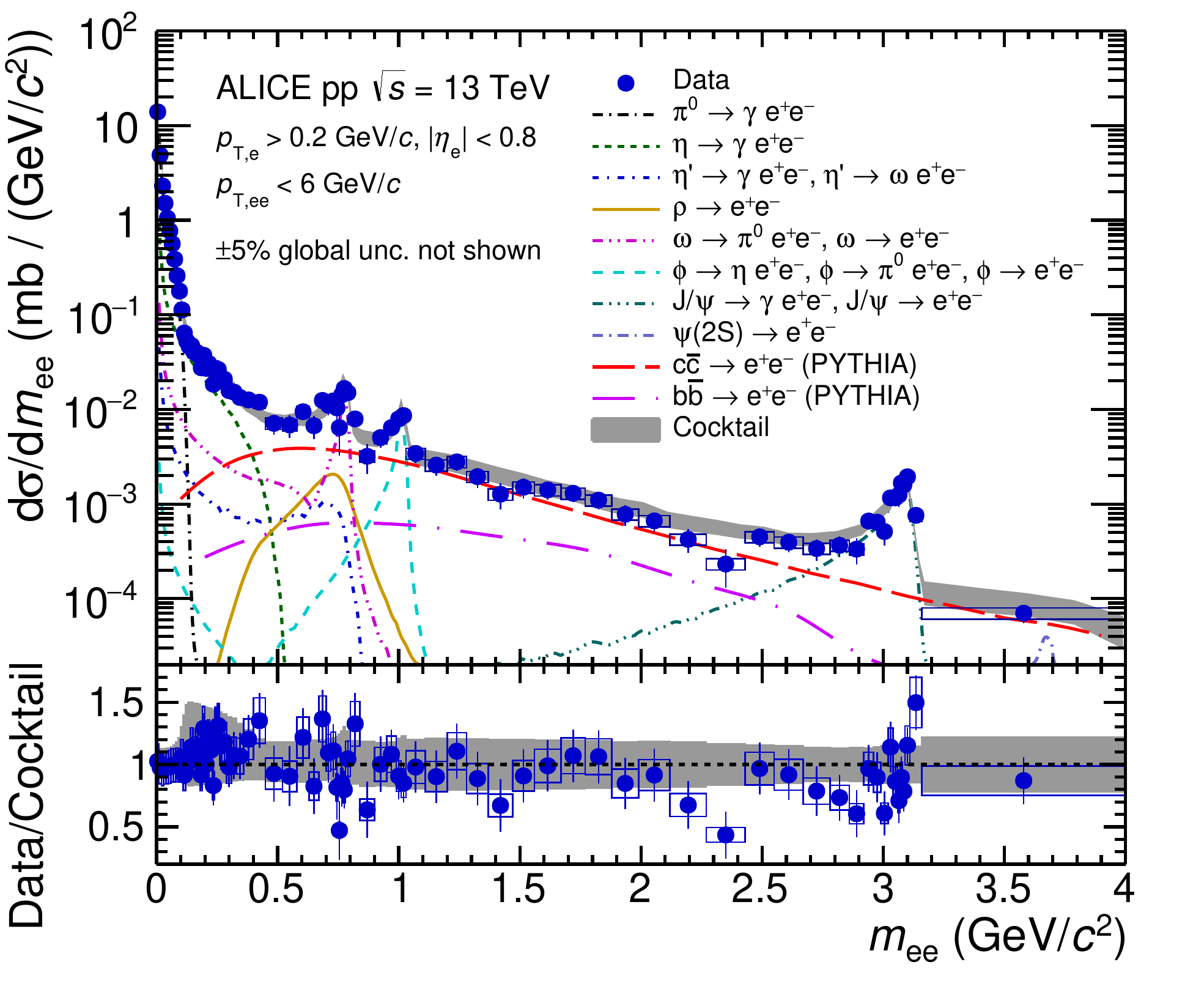}
  \hfill
  \includegraphics[width=0.49\linewidth]{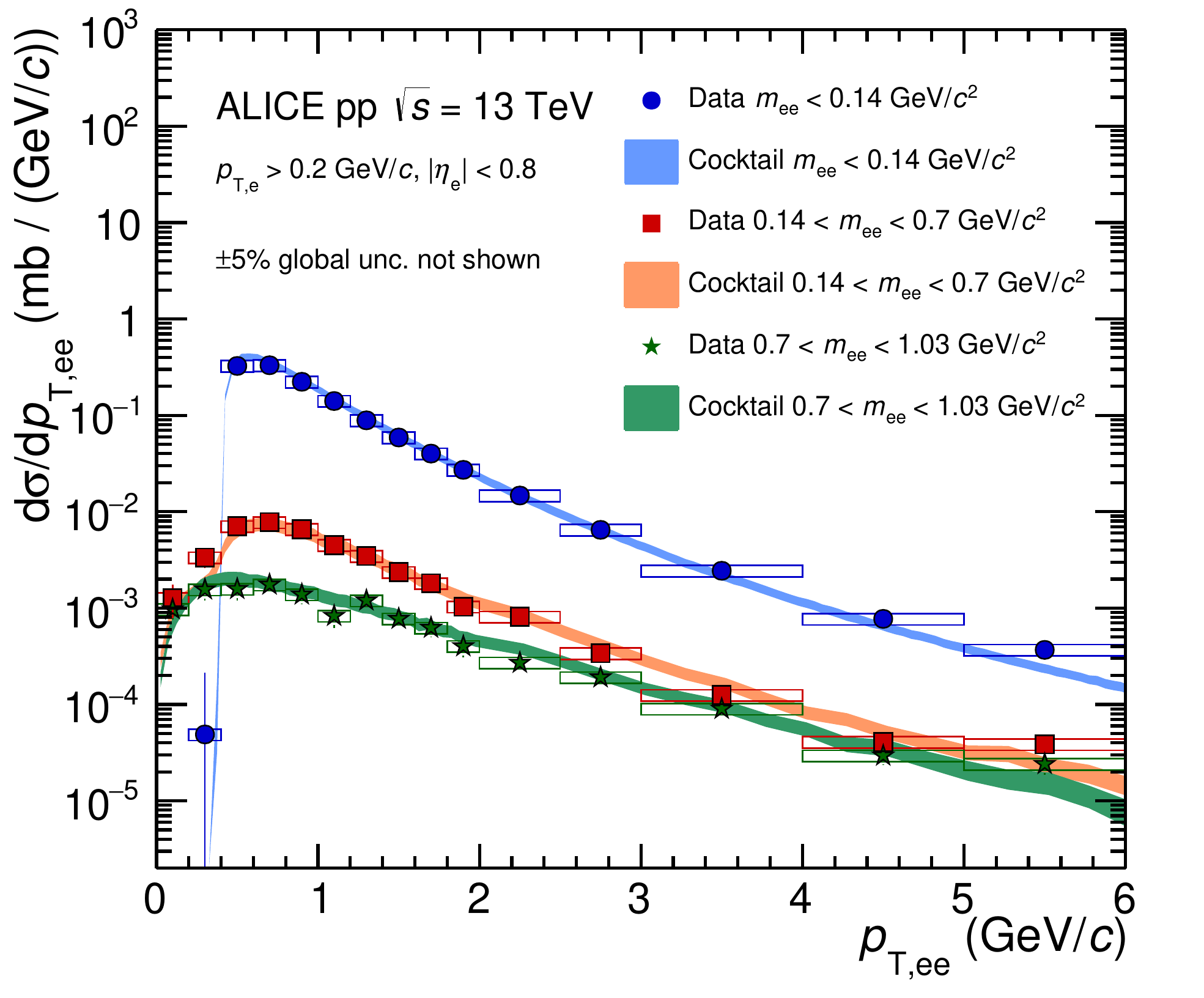}
  \caption{The dielectron cross section in inelastic \pp collisions at
    $\sqrts = 13\TeV$ as a function of invariant mass (left) and of pair
    transverse momentum in different mass intervals (right). The global scale
    uncertainty on the \pp luminosity (5\%) is not shown. The statistical and
    systematic uncertainties of the data are shown as vertical bars and boxes.
    The expectation from the hadronic decay cocktail is shown as a band, and the
    bottom left plot shows the ratio data to cocktail together with the cocktail
    uncertainty.}
  \label{fig:SignalCocktail}
\end{figure}

The dielectron cross sections are reported within the ALICE central barrel
acceptance $\vert\eta_{\rm e}\vert < 0.8$ and $\pte > 0.2\GeVc$, \ie\ without
correction to full phase space. The result, integrated over $\ptee < 6\GeVc$, is
shown as a function of \mee in the left panel of Fig.~\ref{fig:SignalCocktail}.
The data are compared with the expectation from the hadronic decay cocktail,
using \PYTHIA for the heavy-flavour components, and found to be in agreement
within uncertainties. Good agreement between data and cocktail calculations is
also found as a function of \ptee, which is shown for three \mee intervals in
the right panel of Fig.~\ref{fig:SignalCocktail}.

\begin{figure}[ht]
  \centering
  \includegraphics[width=0.49\linewidth]{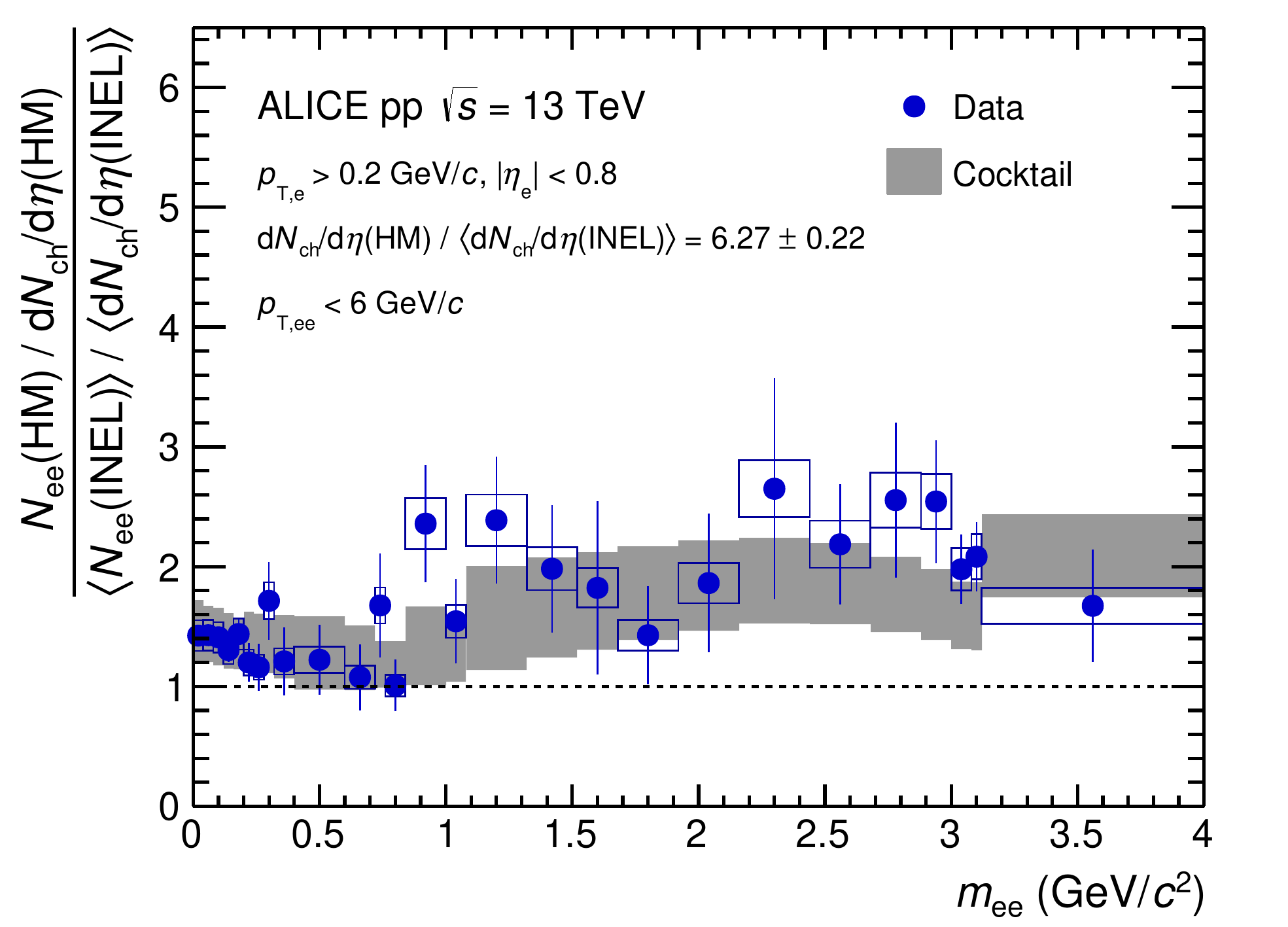}
  \caption{Ratio of dielectron spectra in HM and INEL events scaled by the
    charged-particle multiplicity. The statistical and systematic uncertainties
    of the data are shown as vertical bars and boxes. The expectation from the
    hadronic decay cocktail calculation is shown as a grey band.}
  \label{fig:Ratio_hm_mb_ptint}
\end{figure}

\begin{figure}[ht]
  \centering
  \includegraphics[width=0.49\linewidth]{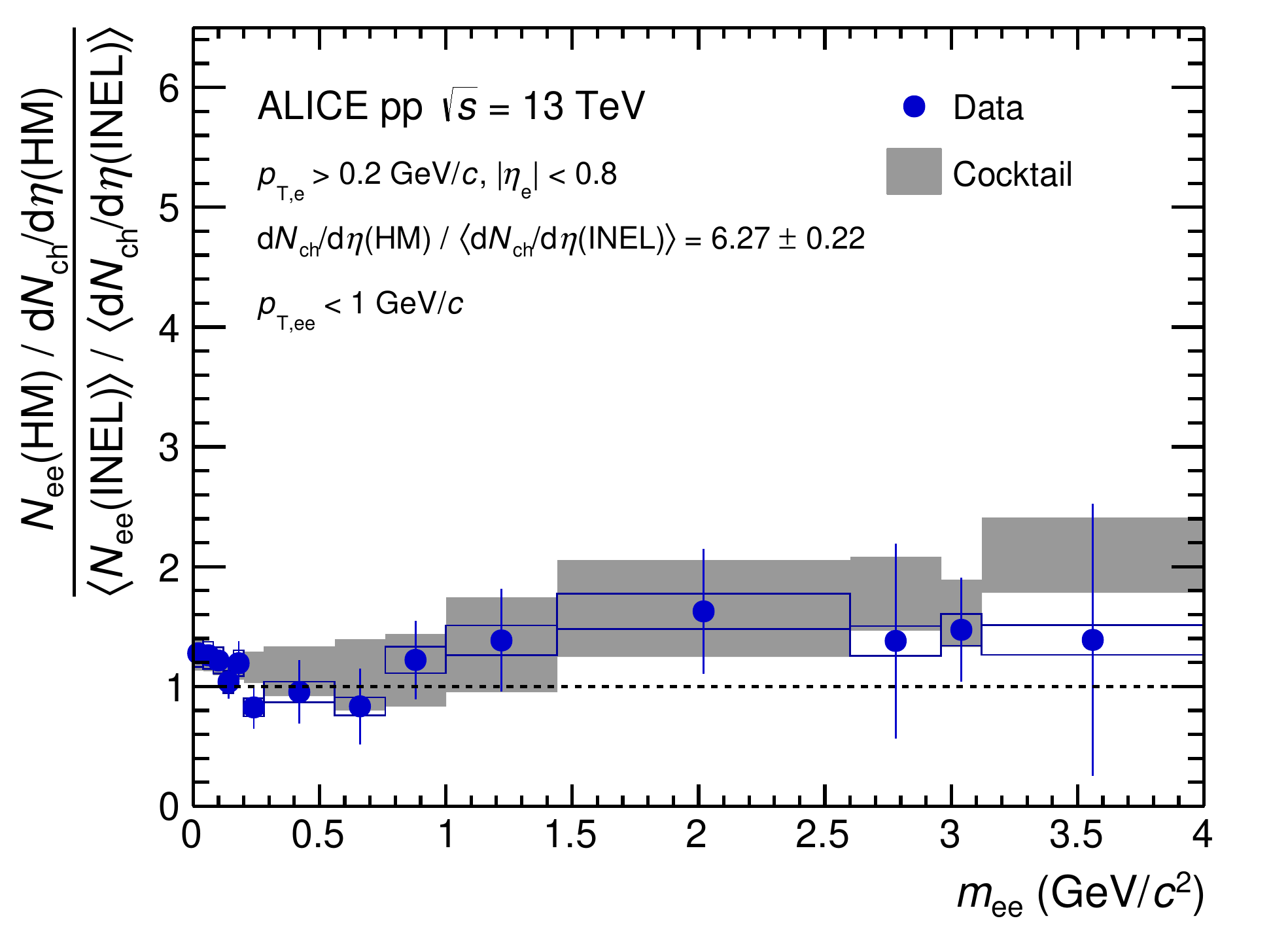}
  \hfill
  \includegraphics[width=0.49\linewidth]{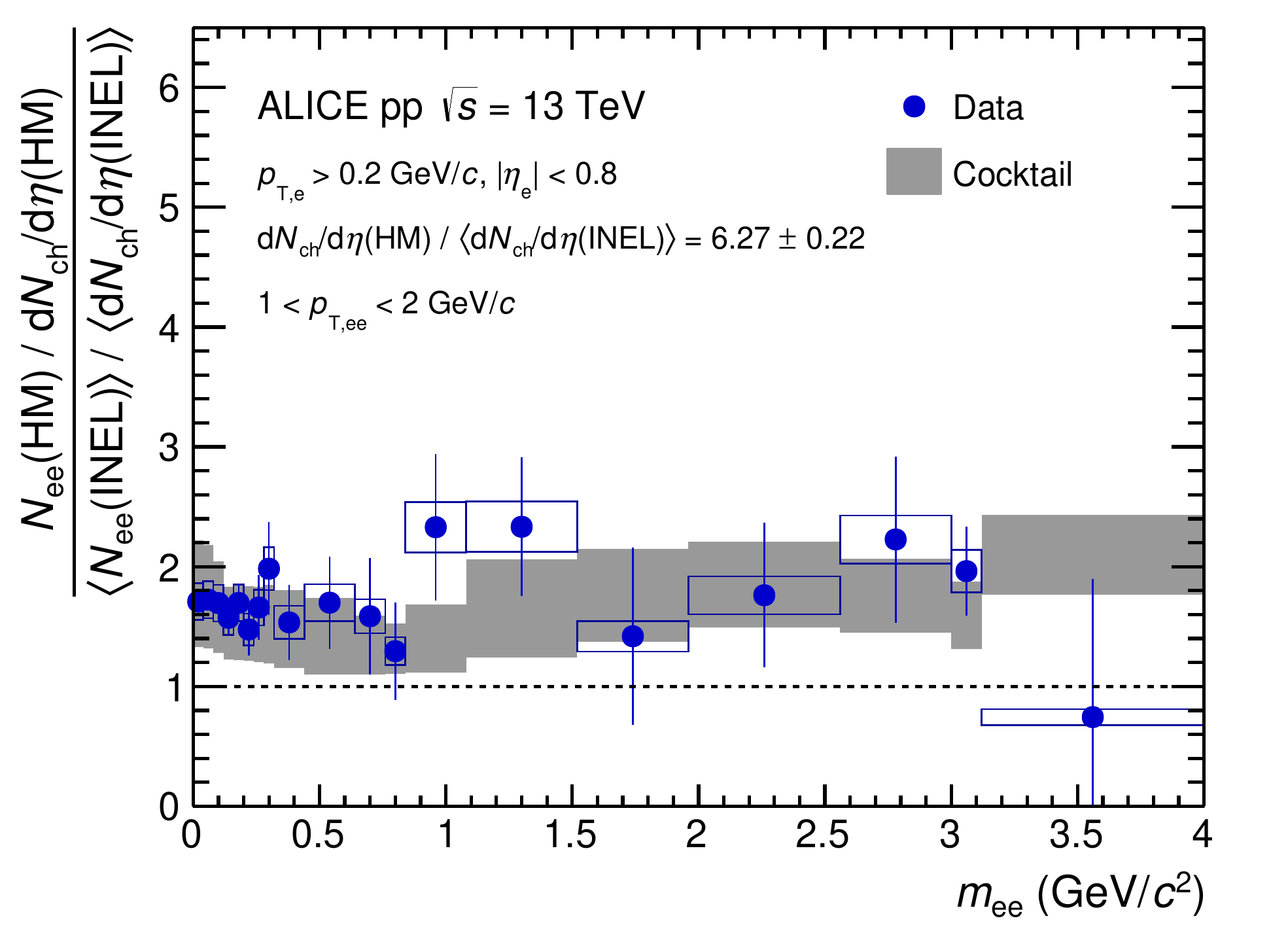}\\
  \includegraphics[width=0.49\linewidth]{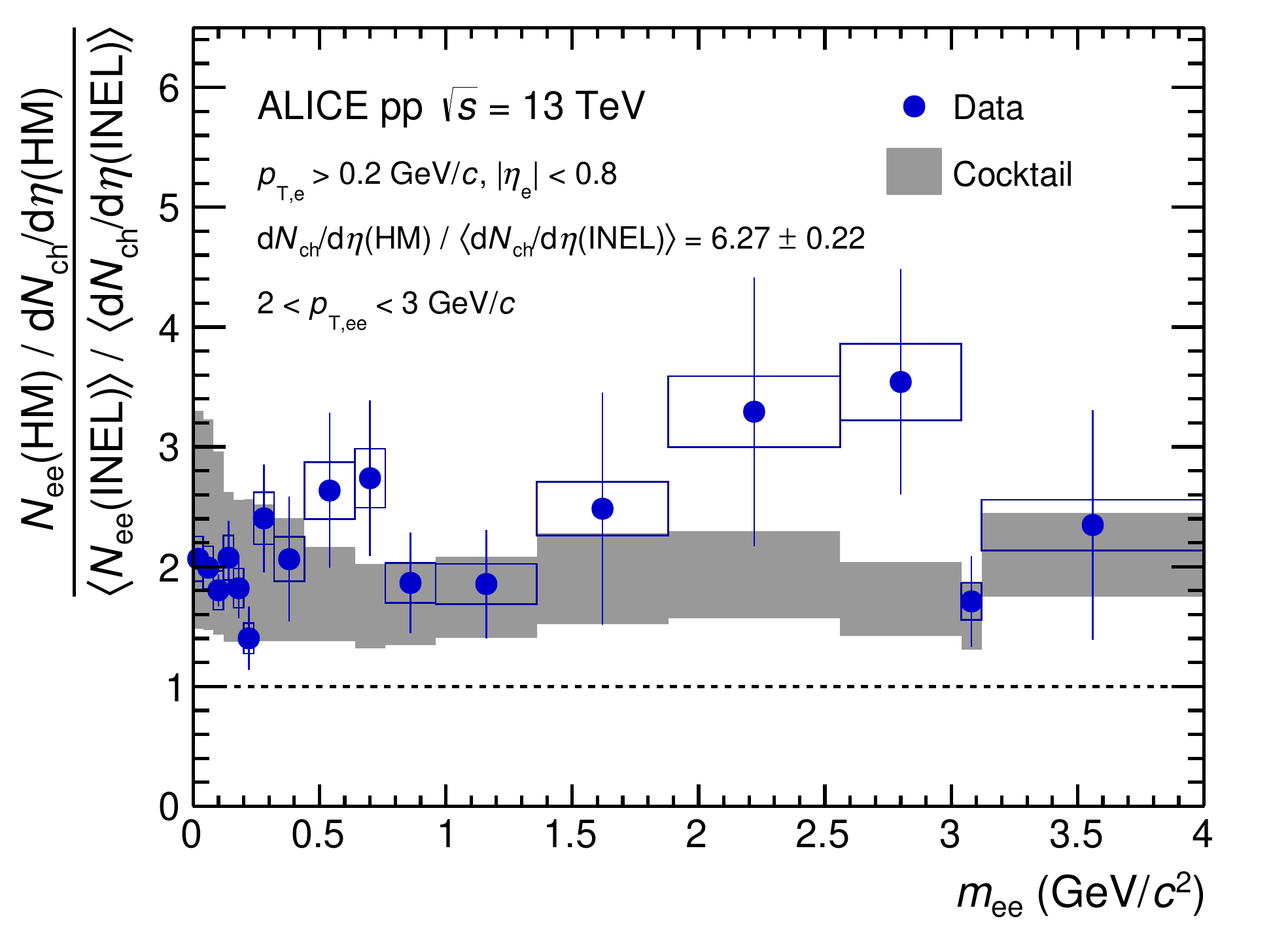}
  \hfill
  \includegraphics[width=0.49\linewidth]{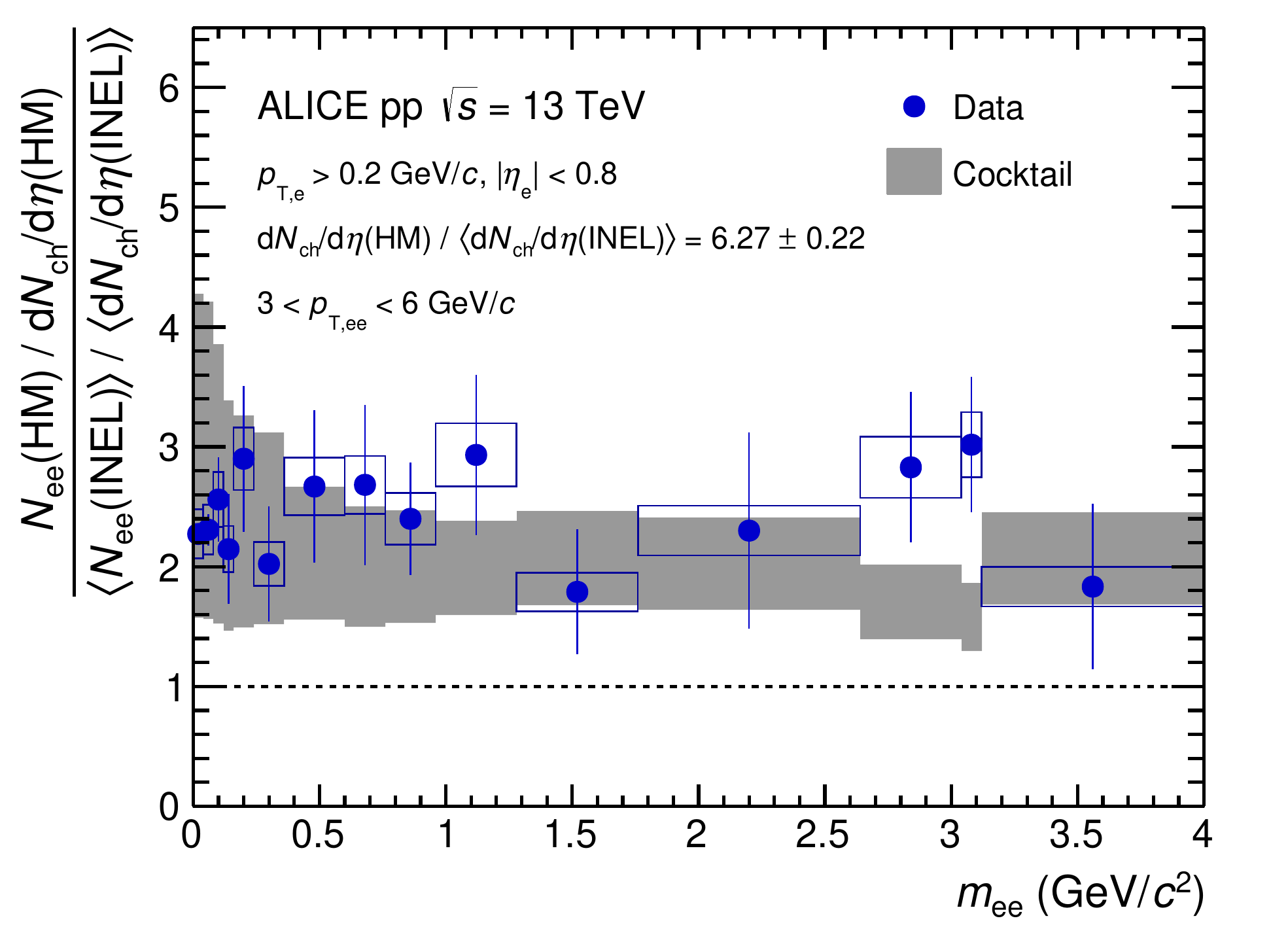}
  \caption{Ratio of dielectron spectra in HM and INEL events scaled by the
    charged-particle multiplicity in different \ptee intervals. The statistical
    and systematic uncertainties of the data are shown as vertical bars and
    boxes. The expectation from the hadronic decay cocktail calculation is shown
    as a grey band.}
  \label{fig:Ratio_hm_mb_ptbins}
\end{figure}

Figures~\ref{fig:Ratio_hm_mb_ptint} and~\ref{fig:Ratio_hm_mb_ptbins} show the
ratios of the dielectron spectra in high-multiplicity over inelastic events as a
function of \mee for different \ptee intervals. To account for the trivial
scaling with charged-particle multiplicity, the ratio is scaled by the factor
$\dd N_{\rm ch}/\dd\eta({\rm HM})/\langle \dd N_{\rm ch}/\dd\eta({\rm
  INEL})\rangle = 6.27 \pm 0.22$, where
$\dd N_{\rm ch}/\dd\eta({\rm HM}) = 33.29 \pm 0.39$ and
$\langle \dd N_{\rm ch}/\dd\eta({\rm INEL})\rangle = 5.31 \pm 0.18$ are the
charged-particle multiplicities in $\vert\eta_{\rm ch}\vert<0.5$ measured in
high-multiplicity and inelastic \pp collisions,
respectively~\cite{Adam:2015pza}. In this ratio, the multiplicity-independent
uncertainties cancel and the total systematic uncertainty reduces to 9\%. The
ratio is in good agreement with the hadronic decay cocktail calculations over
the whole measured \mee and \ptee range. This is the first measurement sensitive
to the production of $\pi^0$, $\eta$, $\omega$ and $\phi$ in high-multiplicity
\pp collisions. The result confirms the hypothesis that these light mesons have
the same multiplicity dependence as a function of \mt, which was used in the
construction of the high-multiplicity hadron cocktail. From the agreement
between data and cocktail in the high-\pt range ($3<\ptee<6\GeVc$), which is
dominated by open beauty, it can be also concluded for the first time that the
open beauty production has a multiplicity dependence similar to that of open
charm. This puts additional constraints on mechanisms used to describe
heavy-flavour production in high-multiplicity \pp collisions, such as multiple
parton interactions, percolation or hydrodynamic models.

\begin{figure}[ht]
  \includegraphics[width=0.49\linewidth]{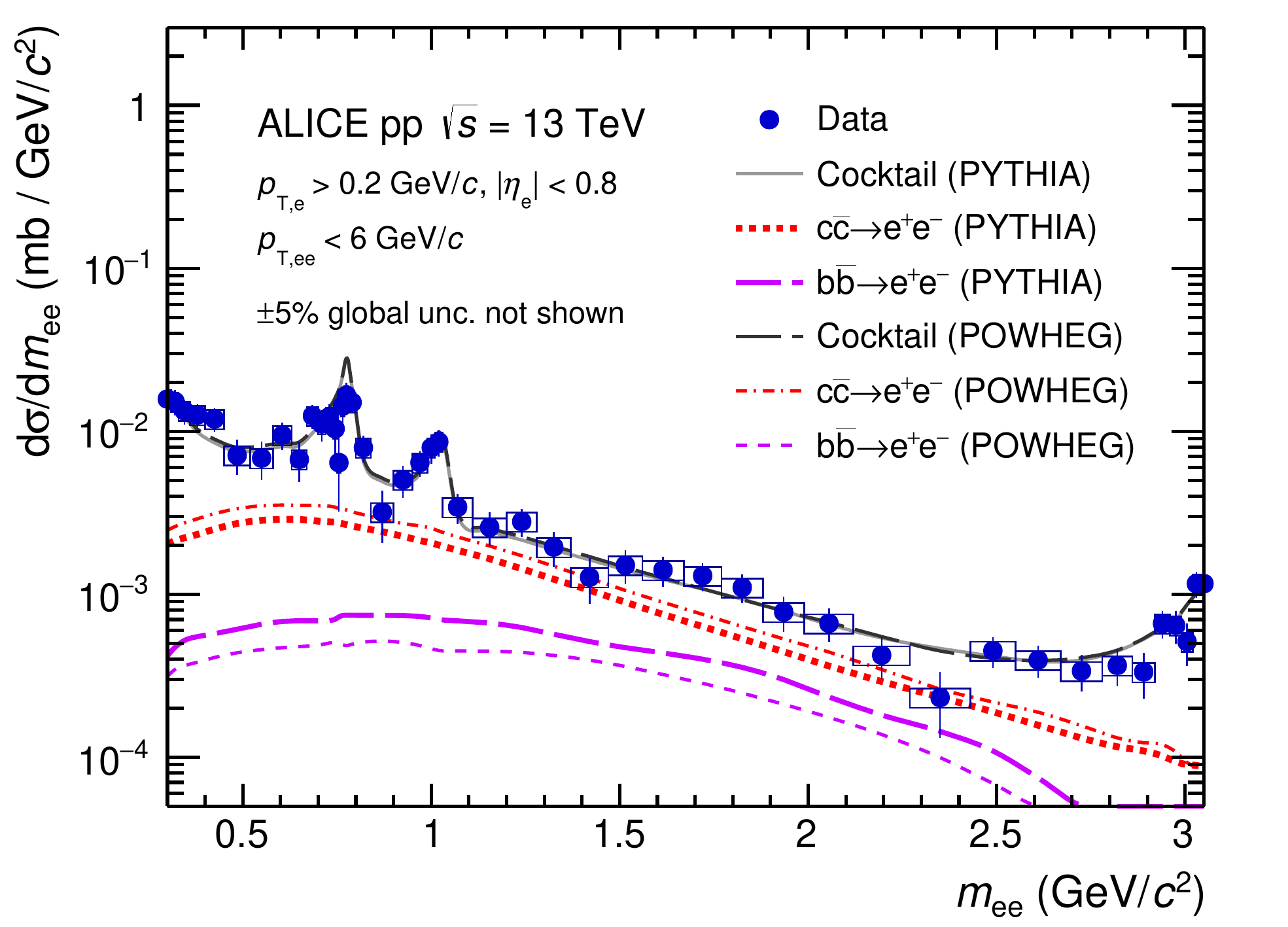}
  \hfill
  \includegraphics[width=0.49\linewidth]{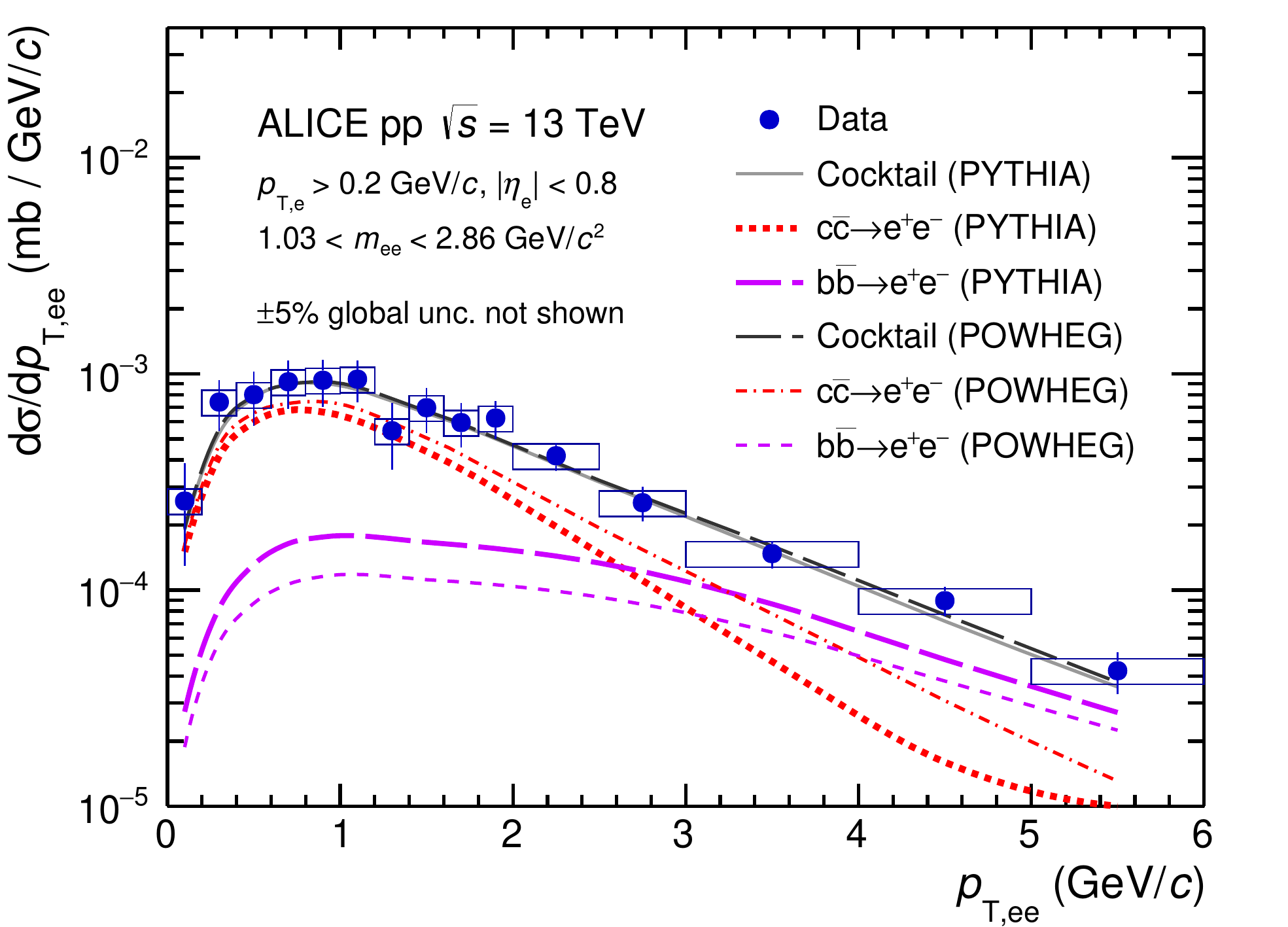}
  \caption{Projection of the heavy-flavour dielectron fit (grey line) in
    inelastic \pp collisions at $\sqrts = 13\TeV$ onto the dielectron mass
    (left) and \ptee (right) using the \PYTHIA and \POWHEG event generators. The
    lines show the charm (red) and beauty (magenta) contributions after the fit.
    The global scale uncertainty on the \pp luminosity (5\%) is not shown. The
    statistical and systematic uncertainties of the data are shown as vertical
    bars and boxes. The fits with \PYTHIA and \POWHEG result in a
    $\chi^2/{\rm ndf}$ of $57.8/66$ and $52.6/66$, respectively.}
  \label{fig:hf_fits_mb}
\end{figure}

In the intermediate mass region ($1.03<\mee<2.86\GeVcc$), which is dominated by
open heavy-flavour decays, the data are fitted simultaneously in \mee and \ptee
(for $\ptee < 6\GeVc$) with \PYTHIA and \POWHEG templates of open charm and
beauty production, keeping the light-flavour and \Jpsi contributions fixed,
which introduces negligible uncertainties on the heavy-flavour cross section.
The \PYTHIA and \POWHEG least-square fits of dielectron spectra in inelastic
events projected over \ptee and \mee are shown in the left and right panels of
Fig.~\ref{fig:hf_fits_mb}, respectively. The resulting cross sections are
summarised in Table~\ref{tab:hfxsection}. The first uncertainty is the
statistical uncertainty resulting from the fits and the second is the systematic
uncertainty, which is determined by moving the data points coherently upward and
downward by their systematic uncertainties and repeating the fits.
The branching fraction of charm-hadron decays to electrons is taken as
$(9.6\pm0.4)\%$~\cite{Patrignani:2016xqp}. An additional uncertainty of of 9.3\%
is added in quadrature to account for differences in the $\Lambda_c/\D^0$ ratio
measured by ALICE in \pp collisions at $\sqrts = 7\TeV$, which is
$0.543\pm0.061\,(\text{stat.})\pm0.160\,(\text{syst})$ for
$\pt > 1\GeVc$~\cite{Acharya:2017kfy}, and the LEP average of
$0.113\pm0.013\pm0.006$~\cite{Gladilin:2014tba}. This translates into a 22\%
uncertainty at the pair level.
The branching fraction of beauty hadrons decaying into electrons, including via
intermediate charm hadrons, is $(21.53\pm0.63)\%$~\cite{Patrignani:2016xqp},
which leads to a 6\% uncertainty on the dielectron-based cross section
measurement. Like the statistical and systematic uncertainties, these branching
fraction uncertainties are fully correlated between the \PYTHIA and \POWHEG
based results.

The results are consistent with extrapolations from lower energies based on pQCD
calculations discussed in the previous section. There is a strong
anti-correlation between the fitted charm and beauty cross sections. The
sizeable difference in the cross sections between the two MC event generators
are comparable to what is observed at $\sqrts = 7\TeV$~\cite{Acharya:2018ohw}.
The different cross sections obtained from fits with \PYTHIA and \POWHEG
simulations are caused by acceptance differences of \ee pairs from heavy-flavour
hadron decays in these two event generators because of different kinematic
correlations of the heavy quark pairs, in particular in rapidity.
The fraction of \ee pairs that fall into the ALICE acceptance and the
intermediate mass region originating from \ccbar pairs at midrapidity is 14\% in
\PYTHIA and 10\% in \POWHEG.
This points to important differences
in the heavy quark production mechanisms between the two generators. It should
be stressed that single heavy-flavour measurements appear insensitive to these
differences as the cross sections obtained from such measurements agree between
\PYTHIA and \POWHEG based
extrapolations~\cite{Aaij:2015bpa,Aad:2015zix,Adam:2016ich}. Therefore,
dielectrons provide complementary information on heavy-flavour production that,
if properly modelled, should lead to consistent cross sections with \PYTHIA and
\POWHEG.

Table~\ref{tab:hfxsection} also summarises the corresponding cross sections for
the high-multiplicity data. In case of \PYTHIA, the measured charm cross section
translates into an enhancement of
$1.86\pm0.40\,(\text{stat.})\pm0.40\,(\text{syst.})$ relative to the
charged-particle multiplicity increase. This is consistent with the modelled
multiplicity dependence used as input for the cocktail in
Figs.~\ref{fig:Ratio_hm_mb_ptint} and~\ref{fig:Ratio_hm_mb_ptbins}. For the
beauty cross section the observed enhancement is
$1.63\pm0.50\,(\text{stat.})\pm0.35\,(\text{syst.})$. This is consistent with
the multiplicity dependence observed for open charm, but a scaling with
charged-particle multiplicity cannot be excluded.

\begin{table}[hb]
  \centering
  \begin{tabular}{l r r}
    \toprule
    & \multicolumn{1}{c}{\PYTHIA} & \multicolumn{1}{c}{\POWHEG} \\
    \midrule
    $\dd\sigma_{\ccbar}/\dd y\vert_{y=0}$ & $\ccPYTHIA \pm \ccPYTHIAstat\,(\text{stat.}) \pm \ccPYTHIAsyst\,(\text{syst.})\mub$ & $\ccPOWHEG \pm \ccPOWHEGstat\,(\text{stat.}) \pm \ccPOWHEGsyst\,(\text{syst.})\mub$ \\\rule{0pt}{3ex}
    $\dd\sigma_{\bbbar}/\dd y\vert_{y=0}$ & $\bbPYTHIA \pm \bbPYTHIAstat\,(\text{stat.}) \pm \bbPYTHIAsyst\,(\text{syst.})\mub$ & $\bbPOWHEG \pm \bbPOWHEGstat\,(\text{stat.}) \pm\,\bbPOWHEGsyst(\text{syst.})\mub$ \\
    \midrule
    $\dd\sigma_{\ccbar}/\dd y\vert_{y=0}^{\rm HM}$ & $\ccHMPYTHIA \pm \ccHMPYTHIAstat\,(\text{stat.}) \pm \ccHMPYTHIAsyst\,(\text{syst.})\mub$ & $\ccHMPOWHEG \pm \ccHMPOWHEGstat\,(\text{stat.}) \pm \ccHMPOWHEGsyst\,(\text{syst.})\mub$ \\\rule{0pt}{3ex}
    $\dd\sigma_{\bbbar}/\dd y\vert_{y=0}^{\rm HM}$ & $\bbHMPYTHIA \pm \bbHMPYTHIAstat\,(\text{stat.}) \pm \bbHMPYTHIAsyst\,(\text{syst.})\mub$ & $\bbHMPOWHEG \pm \bbHMPOWHEGstat\,(\text{stat.}) \pm \bbHMPOWHEGsyst\,(\text{syst.})\mub$ \\
    \bottomrule
  \end{tabular}
  \caption{Heavy-flavour cross sections in inelastic and high-multiplicity \pp
    collisions at $\sqrts = 13\TeV$. The 22\% (6\%) branching fraction
    uncertainty for charm (beauty) decays into electrons is not listed. Like
    statistical and systematic uncertainties, it is fully correlated between the
    \PYTHIA and \POWHEG based results.}
  \label{tab:hfxsection}
\end{table}

The fraction of real direct photons to inclusive photons can be extracted from
the dielectron spectrum at small invariant masses assuming the equivalence
between this fraction and the ratio of virtual direct photons to inclusive
photons. The data are fitted minimising the $\chi^2$, in bins of \ptee, with the
sum of the light-flavour cocktail $\left(f_{\rm LF}(\mee)\right)$, open
heavy-flavour contribution $\left(f_{\rm HF}(\mee)\right)$ and a virtual direct
photon component $\left(f_{\rm direct}(\mee)\right)$, whose shape is described
by the Kroll-Wada equation~\cite{Kroll:1955zu,Landsberg:1986fd} in the
quasi-real virtual photon regime $(\ptee \gg \mee)$. The normalisation of the
open heavy-flavour component is fixed to the measured open charm and beauty
cross sections presented above, using the \PYTHIA simulations for the nominal
fit. As systematic uncertainty estimate, the \POWHEG simulation is used instead.
The light-flavour cocktail and virtual direct photon templates are normalised
independently to the data in $\mee<0.04\GeVcc$, \ie\ in a mass window in which
both Dalitz decays and direct photons have the same $1/\mee$ dependence. The
direct-photon fraction $r$ is then extracted by fitting the data in the mass
interval $0.14<\mee<0.32\GeVcc$, \ie\ above the $\pi^0$ mass to suppress the
most dominant hadron background, with the following expression:
$\dd \sigma/\dd \mee = r f_{\rm dir}(\mee) + (1-r)f_{\rm LF}(\mee) + f_{\rm
  HF}(\mee)$.

No significant direct photon contribution is observed in neither the inelastic
nor the high-multiplicity events~\cite{ALICE-PUBLIC-2018-009}. Upper limits at 90\%
confidence level (C.L.) are extracted with the Feldman--Cousins
method~\cite{Feldman:1997qc} and summarised in Table~\ref{tab:UpperLimits}
together with predictions from perturbative QCD calculations for inelastic
events~\cite{Gordon:1993qc}. The current uncertainties prevent any conclusions
on the scaling of direct-photon production with charged-particle multiplicity.

\begin{table}[ht]
  \centering
  \begin{tabular}{l c c c}
    \toprule
    Data sample & $1 < \ptee < 2\GeVc$ & $2 < \ptee < 3\GeVc$ & $3 < \ptee < 6\GeVc$ \\
    \midrule
    Minimum bias & \DPHonetwo & \DPHtwothree & \DPHthreesix \\
    High multiplicity & \DPHHMonetwo & \DPHHMtwothree & \DPHHMthreesix \\
    \midrule
    pQCD & 0.003 & 0.007 & 0.013 \\
    \bottomrule
  \end{tabular}
  \caption{Upper limits at 90\% C.L. on the direct-photon fractions in
    comparison with the expectation in inelastic \pp collisions based on a NLO
    pQCD calculation for a factorisation and renormalisation scale choice of
    $\mu=\pt$~\cite{Gordon:1993qc}.}
  \label{tab:UpperLimits}
\end{table}

\section{Summary and conclusion}
\label{sec:summary}
We have presented the first measurement of dielectron production at midrapidity
($\vert y_{\mathrm{e}}\vert < 0.8$) in proton--proton collisions at
$\sqrts = 13\TeV$. The dielectron continuum can be well described by the
expected contributions from decays of light- and heavy-flavour hadrons. The
charm and beauty cross sections are extracted for the first time at midrapidity
at $\sqrts = 13\TeV$ and are consistent with extrapolations from lower energies
based on pQCD calculations. The differences observed between \POWHEG and \PYTHIA
imply different kinematic correlations of the heavy-quark pairs in these two
event generators. Therefore dielectrons are uniquely sensitive to the heavy
quark production mechanisms. The comparison of the dielectron spectra in
inelastic events and in events with high charged-particle multiplicities does
not reveal modifications of the spectrum beyond the already established ones of
light and open charm hadrons. The upper limits on the direct-photon fractions
are consistent with predictions from perturbative quantum chromodynamics
calculations.

\newenvironment{acknowledgement}{\relax}{\relax}
\begin{acknowledgement}
\section*{Acknowledgements}
The ALICE Collaboration would like to thank Werner Vogelsang for providing the NLO pQCD calculations for direct
photon production.
\input{fa_2018-05-04.tex}
\end{acknowledgement}

\bibliographystyle{utphys}   
\bibliography{bibliography}

\newpage
\appendix
\section{The ALICE Collaboration}
\label{app:collab}
\input{Alice_Authorlist_2018-May-04.tex}
\end{document}

%% file: fa_2018-05-04.tex
% Version: 2018-05-04

The ALICE Collaboration would like to thank all its engineers and technicians for their invaluable contributions to the construction of the experiment and the CERN accelerator teams for the outstanding performance of the LHC complex.
The ALICE Collaboration gratefully acknowledges the resources and support provided by all Grid centres and the Worldwide LHC Computing Grid (WLCG) collaboration.
The ALICE Collaboration acknowledges the following funding agencies for their support in building and running the ALICE detector:
A. I. Alikhanyan National Science Laboratory (Yerevan Physics Institute) Foundation (ANSL), State Committee of Science and World Federation of Scientists (WFS), Armenia;
Austrian Academy of Sciences and Nationalstiftung f\"{u}r Forschung, Technologie und Entwicklung, Austria;
Ministry of Communications and High Technologies, National Nuclear Research Center, Azerbaijan;
Conselho Nacional de Desenvolvimento Cient\'{\i}fico e Tecnol\'{o}gico (CNPq), Universidade Federal do Rio Grande do Sul (UFRGS), Financiadora de Estudos e Projetos (Finep) and Funda\c{c}\~{a}o de Amparo \`{a} Pesquisa do Estado de S\~{a}o Paulo (FAPESP), Brazil;
Ministry of Science \& Technology of China (MSTC), National Natural Science Foundation of China (NSFC) and Ministry of Education of China (MOEC) , China;
Ministry of Science and Education, Croatia;
Ministry of Education, Youth and Sports of the Czech Republic, Czech Republic;
The Danish Council for Independent Research | Natural Sciences, the Carlsberg Foundation and Danish National Research Foundation (DNRF), Denmark;
Helsinki Institute of Physics (HIP), Finland;
Commissariat \`{a} l'Energie Atomique (CEA) and Institut National de Physique Nucl\'{e}aire et de Physique des Particules (IN2P3) and Centre National de la Recherche Scientifique (CNRS), France;
Bundesministerium f\"{u}r Bildung, Wissenschaft, Forschung und Technologie (BMBF) and GSI Helmholtzzentrum f\"{u}r Schwerionenforschung GmbH, Germany;
General Secretariat for Research and Technology, Ministry of Education, Research and Religions, Greece;
National Research, Development and Innovation Office, Hungary;
Department of Atomic Energy Government of India (DAE), Department of Science and Technology, Government of India (DST), University Grants Commission, Government of India (UGC) and Council of Scientific and Industrial Research (CSIR), India;
Indonesian Institute of Science, Indonesia;
Centro Fermi - Museo Storico della Fisica e Centro Studi e Ricerche Enrico Fermi and Istituto Nazionale di Fisica Nucleare (INFN), Italy;
Institute for Innovative Science and Technology , Nagasaki Institute of Applied Science (IIST), Japan Society for the Promotion of Science (JSPS) KAKENHI and Japanese Ministry of Education, Culture, Sports, Science and Technology (MEXT), Japan;
Consejo Nacional de Ciencia (CONACYT) y Tecnolog\'{i}a, through Fondo de Cooperaci\'{o}n Internacional en Ciencia y Tecnolog\'{i}a (FONCICYT) and Direcci\'{o}n General de Asuntos del Personal Academico (DGAPA), Mexico;
Nederlandse Organisatie voor Wetenschappelijk Onderzoek (NWO), Netherlands;
The Research Council of Norway, Norway;
Commission on Science and Technology for Sustainable Development in the South (COMSATS), Pakistan;
Pontificia Universidad Cat\'{o}lica del Per\'{u}, Peru;
Ministry of Science and Higher Education and National Science Centre, Poland;
Korea Institute of Science and Technology Information and National Research Foundation of Korea (NRF), Republic of Korea;
Ministry of Education and Scientific Research, Institute of Atomic Physics and Romanian National Agency for Science, Technology and Innovation, Romania;
Joint Institute for Nuclear Research (JINR), Ministry of Education and Science of the Russian Federation and National Research Centre Kurchatov Institute, Russia;
Ministry of Education, Science, Research and Sport of the Slovak Republic, Slovakia;
National Research Foundation of South Africa, South Africa;
Centro de Aplicaciones Tecnol\'{o}gicas y Desarrollo Nuclear (CEADEN), Cubaenerg\'{\i}a, Cuba and Centro de Investigaciones Energ\'{e}ticas, Medioambientales y Tecnol\'{o}gicas (CIEMAT), Spain;
Swedish Research Council (VR) and Knut \& Alice Wallenberg Foundation (KAW), Sweden;
European Organization for Nuclear Research, Switzerland;
National Science and Technology Development Agency (NSDTA), Suranaree University of Technology (SUT) and Office of the Higher Education Commission under NRU project of Thailand, Thailand;
Turkish Atomic Energy Agency (TAEK), Turkey;
National Academy of  Sciences of Ukraine, Ukraine;
Science and Technology Facilities Council (STFC), United Kingdom;
National Science Foundation of the United States of America (NSF) and United States Department of Energy, Office of Nuclear Physics (DOE NP), United States of America.

%% file: Alice_Authorlist_2018-May-04.tex
% Collaboration: CERN-LHC-ALICE
% Generation Date is 2018-May-04

% How to use:
%%%%%%%%% appendix with author list
%\appendix
%\section{The ALICE Collaboration}
%\label{app:collab}
%\input{Alice_Authorslist_XXXX-Axx-XX.tex}
\begingroup
\small
\begin{flushleft}
S.~Acharya\Irefn{org139}\And 
F.T.-.~Acosta\Irefn{org20}\And 
D.~Adamov\'{a}\Irefn{org93}\And 
J.~Adolfsson\Irefn{org80}\And 
M.M.~Aggarwal\Irefn{org98}\And 
G.~Aglieri Rinella\Irefn{org34}\And 
M.~Agnello\Irefn{org31}\And 
N.~Agrawal\Irefn{org48}\And 
Z.~Ahammed\Irefn{org139}\And 
S.U.~Ahn\Irefn{org76}\And 
S.~Aiola\Irefn{org144}
% ...
\And
\renewcommand\labelenumi{\textsuperscript{\theenumi}~}

\section*{Affiliation notes}
\renewcommand\theenumi{\roman{enumi}}
\begin{Authlist}
\item \Adef{org*}Deceased
\item \Adef{orgI}Dipartimento DET del Politecnico di Torino, Turin, Italy
\item \Adef{orgII}M.V. Lomonosov Moscow State University, D.V. Skobeltsyn Institute of Nuclear, Physics, Moscow, Russia
\item \Adef{orgIII}Department of Applied Physics, Aligarh Muslim University, Aligarh, India
\item \Adef{orgIV}Institute of Theoretical Physics, University of Wroclaw, Poland
\end{Authlist}

\section*{Collaboration Institutes}
\renewcommand\theenumi{\arabic{enumi}~}
\begin{Authlist}
\item \Idef{org1}A.I. Alikhanyan National Science Laboratory (Yerevan Physics Institute) Foundation, Yerevan, Armenia
\item \Idef{org2}Bogolyubov Institute for Theoretical Physics, National Academy of Sciences of Ukraine, Kiev, Ukraine
\item \Idef{org3}Bose Institute, Department of Physics  and Centre for Astroparticle Physics and Space Science (CAPSS), Kolkata, India
\item \Idef{org4}Budker Institute for Nuclear Physics, Novosibirsk, Russia
\item \Idef{org5}California Polytechnic State University, San Luis Obispo, California, United States
\item \Idef{org6}Central China Normal University, Wuhan, China
\item \Idef{org7}Centre de Calcul de l'IN2P3, Villeurbanne, Lyon, France
\item \Idef{org8}Centro de Aplicaciones Tecnol\'{o}gicas y Desarrollo Nuclear (CEADEN), Havana, Cuba
\item \Idef{org9}Centro de Investigaci\'{o}n y de Estudios Avanzados (CINVESTAV), Mexico City and M\'{e}rida, Mexico
\item \Idef{org10}Centro Fermi - Museo Storico della Fisica e Centro Studi e Ricerche ``Enrico Fermi', Rome, Italy
\item \Idef{org11}Chicago State University, Chicago, Illinois, United States
\item \Idef{org12}China Institute of Atomic Energy, Beijing, China
\item \Idef{org13}Chonbuk National University, Jeonju, Republic of Korea
\item \Idef{org14}Comenius University Bratislava, Faculty of Mathematics, Physics and Informatics, Bratislava, Slovakia
\item \Idef{org15}COMSATS Institute of Information Technology (CIIT), Islamabad, Pakistan
\item \Idef{org16}Creighton University, Omaha, Nebraska, United States
\item \Idef{org17}Department of Physics, Aligarh Muslim University, Aligarh, India
\item \Idef{org18}Department of Physics, Pusan National University, Pusan, Republic of Korea
\item \Idef{org19}Department of Physics, Sejong University, Seoul, Republic of Korea
\item \Idef{org20}Department of Physics, University of California, Berkeley, California, United States
\item \Idef{org21}Department of Physics, University of Oslo, Oslo, Norway
\item \Idef{org22}Department of Physics and Technology, University of Bergen, Bergen, Norway
\item \Idef{org23}Dipartimento di Fisica dell'Universit\`{a} 'La Sapienza' and Sezione INFN, Rome, Italy
\item \Idef{org24}Dipartimento di Fisica dell'Universit\`{a} and Sezione INFN, Cagliari, Italy
\item \Idef{org25}Dipartimento di Fisica dell'Universit\`{a} and Sezione INFN, Trieste, Italy
\item \Idef{org26}Dipartimento di Fisica dell'Universit\`{a} and Sezione INFN, Turin, Italy
\item \Idef{org27}Dipartimento di Fisica e Astronomia dell'Universit\`{a} and Sezione INFN, Bologna, Italy
\item \Idef{org28}Dipartimento di Fisica e Astronomia dell'Universit\`{a} and Sezione INFN, Catania, Italy
\item \Idef{org29}Dipartimento di Fisica e Astronomia dell'Universit\`{a} and Sezione INFN, Padova, Italy
\item \Idef{org30}Dipartimento di Fisica `E.R.~Caianiello' dell'Universit\`{a} and Gruppo Collegato INFN, Salerno, Italy
\item \Idef{org31}Dipartimento DISAT del Politecnico and Sezione INFN, Turin, Italy
\item \Idef{org32}Dipartimento di Scienze e Innovazione Tecnologica dell'Universit\`{a} del Piemonte Orientale and INFN Sezione di Torino, Alessandria, Italy
\item \Idef{org33}Dipartimento Interateneo di Fisica `M.~Merlin' and Sezione INFN, Bari, Italy
\item \Idef{org34}European Organization for Nuclear Research (CERN), Geneva, Switzerland
\item \Idef{org35}Faculty of Electrical Engineering, Mechanical Engineering and Naval Architecture, University of Split, Split, Croatia
\item \Idef{org36}Faculty of Engineering and Science, Western Norway University of Applied Sciences, Bergen, Norway
\item \Idef{org37}Faculty of Nuclear Sciences and Physical Engineering, Czech Technical University in Prague, Prague, Czech Republic
\item \Idef{org38}Faculty of Science, P.J.~\v{S}af\'{a}rik University, Ko\v{s}ice, Slovakia
\item \Idef{org39}Frankfurt Institute for Advanced Studies, Johann Wolfgang Goethe-Universit\"{a}t Frankfurt, Frankfurt, Germany
\item \Idef{org40}Gangneung-Wonju National University, Gangneung, Republic of Korea
\item \Idef{org41}Gauhati University, Department of Physics, Guwahati, India
\item \Idef{org42}Helmholtz-Institut f\"{u}r Strahlen- und Kernphysik, Rheinische Friedrich-Wilhelms-Universit\"{a}t Bonn, Bonn, Germany
\item \Idef{org43}Helsinki Institute of Physics (HIP), Helsinki, Finland
\item \Idef{org44}High Energy Physics Group,  Universidad Aut\'{o}noma de Puebla, Puebla, Mexico
\item \Idef{org45}Hiroshima University, Hiroshima, Japan
\item \Idef{org46}Hochschule Worms, Zentrum  f\"{u}r Technologietransfer und Telekommunikation (ZTT), Worms, Germany
\item \Idef{org47}Horia Hulubei National Institute of Physics and Nuclear Engineering, Bucharest, Romania
\item \Idef{org48}Indian Institute of Technology Bombay (IIT), Mumbai, India
\item \Idef{org49}Indian Institute of Technology Indore, Indore, India
\item \Idef{org50}Indonesian Institute of Sciences, Jakarta, Indonesia
\item \Idef{org51}INFN, Laboratori Nazionali di Frascati, Frascati, Italy
\item \Idef{org52}INFN, Sezione di Bari, Bari, Italy
\item \Idef{org53}INFN, Sezione di Bologna, Bologna, Italy
\item \Idef{org54}INFN, Sezione di Cagliari, Cagliari, Italy
\item \Idef{org55}INFN, Sezione di Catania, Catania, Italy
\item \Idef{org56}INFN, Sezione di Padova, Padova, Italy
\item \Idef{org57}INFN, Sezione di Roma, Rome, Italy
\item \Idef{org58}INFN, Sezione di Torino, Turin, Italy
\item \Idef{org59}INFN, Sezione di Trieste, Trieste, Italy
\item \Idef{org60}Inha University, Incheon, Republic of Korea
\item \Idef{org61}Institut de Physique Nucl\'{e}aire d'Orsay (IPNO), Institut National de Physique Nucl\'{e}aire et de Physique des Particules (IN2P3/CNRS), Universit\'{e} de Paris-Sud, Universit\'{e} Paris-Saclay, Orsay, France
\item \Idef{org62}Institute for Nuclear Research, Academy of Sciences, Moscow, Russia
\item \Idef{org63}Institute for Subatomic Physics, Utrecht University/Nikhef, Utrecht, Netherlands
\item \Idef{org64}Institute for Theoretical and Experimental Physics, Moscow, Russia
\item \Idef{org65}Institute of Experimental Physics, Slovak Academy of Sciences, Ko\v{s}ice, Slovakia
\item \Idef{org66}Institute of Physics, Homi Bhabha National Institute, Bhubaneswar, India
\item \Idef{org67}Institute of Physics of the Czech Academy of Sciences, Prague, Czech Republic
\item \Idef{org68}Institute of Space Science (ISS), Bucharest, Romania
\item \Idef{org69}Institut f\"{u}r Kernphysik, Johann Wolfgang Goethe-Universit\"{a}t Frankfurt, Frankfurt, Germany
\item \Idef{org70}Instituto de Ciencias Nucleares, Universidad Nacional Aut\'{o}noma de M\'{e}xico, Mexico City, Mexico
\item \Idef{org71}Instituto de F\'{i}sica, Universidade Federal do Rio Grande do Sul (UFRGS), Porto Alegre, Brazil
\item \Idef{org72}Instituto de F\'{\i}sica, Universidad Nacional Aut\'{o}noma de M\'{e}xico, Mexico City, Mexico
\item \Idef{org73}iThemba LABS, National Research Foundation, Somerset West, South Africa
\item \Idef{org74}Johann-Wolfgang-Goethe Universit\"{a}t Frankfurt Institut f\"{u}r Informatik, Fachbereich Informatik und Mathematik, Frankfurt, Germany
\item \Idef{org75}Joint Institute for Nuclear Research (JINR), Dubna, Russia
\item \Idef{org76}Korea Institute of Science and Technology Information, Daejeon, Republic of Korea
\item \Idef{org77}KTO Karatay University, Konya, Turkey
\item \Idef{org78}Laboratoire de Physique Subatomique et de Cosmologie, Universit\'{e} Grenoble-Alpes, CNRS-IN2P3, Grenoble, France
\item \Idef{org79}Lawrence Berkeley National Laboratory, Berkeley, California, United States
\item \Idef{org80}Lund University Department of Physics, Division of Particle Physics, Lund, Sweden
\item \Idef{org81}Nagasaki Institute of Applied Science, Nagasaki, Japan
\item \Idef{org82}Nara Women{'}s University (NWU), Nara, Japan
\item \Idef{org83}National and Kapodistrian University of Athens, School of Science, Department of Physics , Athens, Greece
\item \Idef{org84}National Centre for Nuclear Research, Warsaw, Poland
\item \Idef{org85}National Institute of Science Education and Research, Homi Bhabha National Institute, Jatni, India
\item \Idef{org86}National Nuclear Research Center, Baku, Azerbaijan
\item \Idef{org87}National Research Centre Kurchatov Institute, Moscow, Russia
\item \Idef{org88}Niels Bohr Institute, University of Copenhagen, Copenhagen, Denmark
\item \Idef{org89}Nikhef, National institute for subatomic physics, Amsterdam, Netherlands
\item \Idef{org90}NRC Kurchatov Institute IHEP, Protvino, Russia
\item \Idef{org91}NRNU Moscow Engineering Physics Institute, Moscow, Russia
\item \Idef{org92}Nuclear Physics Group, STFC Daresbury Laboratory, Daresbury, United Kingdom
\item \Idef{org93}Nuclear Physics Institute of the Czech Academy of Sciences, \v{R}e\v{z} u Prahy, Czech Republic
\item \Idef{org94}Oak Ridge National Laboratory, Oak Ridge, Tennessee, United States
\item \Idef{org95}Ohio State University, Columbus, Ohio, United States
\item \Idef{org96}Petersburg Nuclear Physics Institute, Gatchina, Russia
\item \Idef{org97}Physics department, Faculty of science, University of Zagreb, Zagreb, Croatia
\item \Idef{org98}Physics Department, Panjab University, Chandigarh, India
\item \Idef{org99}Physics Department, University of Jammu, Jammu, India
\item \Idef{org100}Physics Department, University of Rajasthan, Jaipur, India
\item \Idef{org101}Physikalisches Institut, Eberhard-Karls-Universit\"{a}t T\"{u}bingen, T\"{u}bingen, Germany
\item \Idef{org102}Physikalisches Institut, Ruprecht-Karls-Universit\"{a}t Heidelberg, Heidelberg, Germany
\item \Idef{org103}Physik Department, Technische Universit\"{a}t M\"{u}nchen, Munich, Germany
\item \Idef{org104}Research Division and ExtreMe Matter Institute EMMI, GSI Helmholtzzentrum f\"ur Schwerionenforschung GmbH, Darmstadt, Germany
\item \Idef{org105}Rudjer Bo\v{s}kovi\'{c} Institute, Zagreb, Croatia
\item \Idef{org106}Russian Federal Nuclear Center (VNIIEF), Sarov, Russia
\item \Idef{org107}Saha Institute of Nuclear Physics, Homi Bhabha National Institute, Kolkata, India
\item \Idef{org108}School of Physics and Astronomy, University of Birmingham, Birmingham, United Kingdom
\item \Idef{org109}Secci\'{o}n F\'{\i}sica, Departamento de Ciencias, Pontificia Universidad Cat\'{o}lica del Per\'{u}, Lima, Peru
\item \Idef{org110}Shanghai Institute of Applied Physics, Shanghai, China
\item \Idef{org111}St. Petersburg State University, St. Petersburg, Russia
\item \Idef{org112}Stefan Meyer Institut f\"{u}r Subatomare Physik (SMI), Vienna, Austria
\item \Idef{org113}SUBATECH, IMT Atlantique, Universit\'{e} de Nantes, CNRS-IN2P3, Nantes, France
\item \Idef{org114}Suranaree University of Technology, Nakhon Ratchasima, Thailand
\item \Idef{org115}Technical University of Ko\v{s}ice, Ko\v{s}ice, Slovakia
\item \Idef{org116}Technische Universit\"{a}t M\"{u}nchen, Excellence Cluster 'Universe', Munich, Germany
\item \Idef{org117}The Henryk Niewodniczanski Institute of Nuclear Physics, Polish Academy of Sciences, Cracow, Poland
\item \Idef{org118}The University of Texas at Austin, Austin, Texas, United States
\item \Idef{org119}Universidad Aut\'{o}noma de Sinaloa, Culiac\'{a}n, Mexico
\item \Idef{org120}Universidade de S\~{a}o Paulo (USP), S\~{a}o Paulo, Brazil
\item \Idef{org121}Universidade Estadual de Campinas (UNICAMP), Campinas, Brazil
\item \Idef{org122}Universidade Federal do ABC, Santo Andre, Brazil
\item \Idef{org123}University College of Southeast Norway, Tonsberg, Norway
\item \Idef{org124}University of Cape Town, Cape Town, South Africa
\item \Idef{org125}University of Houston, Houston, Texas, United States
\item \Idef{org126}University of Jyv\"{a}skyl\"{a}, Jyv\"{a}skyl\"{a}, Finland
\item \Idef{org127}University of Liverpool, Liverpool, United Kingdom
\item \Idef{org128}University of Tennessee, Knoxville, Tennessee, United States
\item \Idef{org129}University of the Witwatersrand, Johannesburg, South Africa
\item \Idef{org130}University of Tokyo, Tokyo, Japan
\item \Idef{org131}University of Tsukuba, Tsukuba, Japan
\item \Idef{org132}Universit\'{e} Clermont Auvergne, CNRS/IN2P3, LPC, Clermont-Ferrand, France
\item \Idef{org133}Universit\'{e} de Lyon, Universit\'{e} Lyon 1, CNRS/IN2P3, IPN-Lyon, Villeurbanne, Lyon, France
\item \Idef{org134}Universit\'{e} de Strasbourg, CNRS, IPHC UMR 7178, F-67000 Strasbourg, France, Strasbourg, France
\item \Idef{org135} Universit\'{e} Paris-Saclay Centre d¿\'Etudes de Saclay (CEA), IRFU, Department de Physique Nucl\'{e}aire (DPhN), Saclay, France
\item \Idef{org136}Universit\`{a} degli Studi di Foggia, Foggia, Italy
\item \Idef{org137}Universit\`{a} degli Studi di Pavia, Pavia, Italy
\item \Idef{org138}Universit\`{a} di Brescia, Brescia, Italy
\item \Idef{org139}Variable Energy Cyclotron Centre, Homi Bhabha National Institute, Kolkata, India
\item \Idef{org140}Warsaw University of Technology, Warsaw, Poland
\item \Idef{org141}Wayne State University, Detroit, Michigan, United States
\item \Idef{org142}Westf\"{a}lische Wilhelms-Universit\"{a}t M\"{u}nster, Institut f\"{u}r Kernphysik, M\"{u}nster, Germany
\item \Idef{org143}Wigner Research Centre for Physics, Hungarian Academy of Sciences, Budapest, Hungary
\item \Idef{org144}Yale University, New Haven, Connecticut, United States
\item \Idef{org145}Yonsei University, Seoul, Republic of Korea
\end{Authlist}
\endgroup